\documentclass[journal]{IEEEtran}
\usepackage{amsmath, cleveref}

\usepackage{multirow}
\usepackage{wrapfig}
\usepackage{subfigure}
\usepackage{mathtools}
\usepackage{color}
\usepackage{rotating}
\usepackage{cite}
\usepackage{array,url}
\usepackage{balance}
\usepackage{tabularx}
\usepackage{epstopdf}
\def \etal {et al.}

\ifCLASSINFOpdf
\else
\fi
\begin{document}
%
\title{Speech-Driven Animation with Meaningful Behaviors}
%
%
%

\author{Najmeh~Sadoughi,~\IEEEmembership{Student Member,~IEEE,}
        and Carlos~Busso,~\IEEEmembership{Senior Member,~IEEE}
\thanks{N. Sadoughi and C. Busso are with the Department
of Electrical and Computer Engineering, The university of Texas at Dallas, Dallas,
TX, 75080 USA (e-mail: nxs137130@utdallas.edu and busso@utdallas.edu).}
}

%
%

\markboth{}%
{Shell \MakeLowercase{\textit{\etal}}: Bare Demo of IEEEtran.cls for Journals}
%



\maketitle

\begin{abstract}
\emph{Conversational agents} (CAs) play an important role in \emph{human computer interaction} (HCI). Creating believable movements for CAs is challenging, since the movements have to be meaningful and natural, reflecting the coupling between gestures and speech.  Studies in the past have mainly relied on rule-based or data-driven approaches. Rule-based methods focus on creating meaningful behaviors conveying the underlying message, but the gestures cannot be easily synchronized with speech. Data-driven approaches, especially speech-driven models, can capture the relationship between speech and gestures. However, they create behaviors disregarding the meaning of the message. This study proposes to bridge the gap between these two approaches overcoming their limitations. The approach builds a \emph{dynamic Bayesian network} (DBN), where a discrete variable is added to constrain the behaviors on the underlying constraint. The study implements and evaluates the approach with two constraints: discourse functions and prototypical behaviors. By constraining on the discourse functions (e.g., questions), the model learns the characteristic behaviors associated with a given discourse class learning the rules from the data. By constraining on prototypical behaviors (e.g., head nods), the approach can be embedded in a rule-based system as a behavior realizer creating trajectories that are timely synchronized with speech. The study proposes a DBN structure and a training approach that (1) models the cause-effect relationship between the constraint and the gestures, (2) initializes the state configuration models increasing the range of the generated behaviors, and (3) captures the differences in the behaviors across constraints by enforcing sparse transitions between shared and exclusive states per constraint. Objective and subjective evaluations demonstrate the benefits of the proposed approach over an unconstrained baseline model.
\end{abstract}

\begin{IEEEkeywords}
Speech-driven animation, Conversation agents, Behavior Synthesis, Discourse function.
\end{IEEEkeywords}

%
\IEEEpeerreviewmaketitle

\section{Introduction}
\label{sec:intro}


Body language is an essential part of face-to-face conversations. People consciously or unconsciously use head motion, hand gestures, and facial expressions while speaking. We use these modalities for multiple purposes including to emphasize ideas, parse sentences into smaller syntactic units, complement verbal information, and express our emotions. Therefore, incorporating naturalistic behaviors that fulfill these communication goals is important in the design of a \emph{conversational agent} (CA) \cite{Cassell_1994}.  CAs are playing a relevant role in several fields including business enterprises, healthcare, entertainment, and education. Their use has also increased with new website and mobile applications, providing a great platform for virtual reality, visual aid for hearing impaired individuals, and virtual agents for online shopping \cite{Chattaraman_2014}. 

Creating behaviors that are perceived natural while conveying the underlying meaning in the message is challenging. Most studies in this field are focused on either rule-based or data-driven systems \cite{Sadoughi_2017_2}. Rule-based systems create contextual rules to trigger behaviors, emphasizing the semantic and syntactic information \cite{Cassell_1994, Kopp_2006, Bevacqua_2007}. However, the variation of the gestures generated using rules is bounded by the predefined dictionary of handmade gestures \cite{Foster_2007}. Furthermore, scheduling the movement with speech is challenging \cite{Welbergen_2005, Cassell_2004}. Data-driven approaches learn the behaviors directly from data. There are several studies that have used speech to create behaviors \cite{Brand_1999, Cao_2005, Busso_2007, Mariooryad_2012_2, Levine_2010, Chiu_2011, Le_2012}. Speech prosody is highly correlated with facial expression and head movements  \cite{Busso_2007_3}, so it is possible to generate behaviors that are timely aligned with speech (rhythm, emphasis). However, speech-driven methods do not consider the meaning of the sentence. While the gesture may be perfectly aligned with speech, its meaning may contradict the message (e.g., nodding while saying ``no"). 

This study leverages the advantages of rule-based and speech-driven systems, bridging the gap between these methods by overcoming their drawbacks. We address this problem by constraining our speech-driven model by contextual information to generate behaviors with meaning. This approach relies on \emph{dynamic Bayesian models} (DBNs) capturing the temporal relationship between speech and gestures (in this study, hand and head motion). We introduce the constraints as an extra discrete variable that conditions the state configuration between speech and gestures, modeling the specific behavioral characteristics associated with the given constraint. We demonstrate the potential of the proposed approach with two evaluations, where the constraints are either discourse functions (negation, affirmation, questions and suggestion) or predefined prototypical hand and head gestures. With the discourse functions, we aim to synthesize head movement trajectories that are commonly associated with a given discourse function (e.g., head roll for questions, head shakes for negations). Instead of creating hand-crafted rules, the proposed model learns the statistical patterns from data. For prototypical gestures, we aim to learn statistical models that generate pre-defined hand and head behaviors, and their joint representations with prosody features. This model plays the role of a behavior realizer in the SAIBA framework \cite{Kopp_2006}, and has the potential to be integrated into a rule-based system. We consider three prototypical hand gestures (\emph{To-Fro}, \emph{So-What}, and \emph{Regress}) and head gesture (\emph{Head Nod} and \emph{Head Shake}). The constraints are introduced as input to the system, changing the discrete variable that conditions the generated gestures. During synthesis, the models will create novel realizations of these gestures that are timely synchronized with speech. The proposed models are effective, producing realizations that are perceived more natural than the unconstrained models, bridging the gap between rule-based and speech-driven methods. 


\section{Related Work}
\label{sec:litrev}

Several studies have proposed schemes to  generate gestures, which can be categorized into rule-based and data-driven methods.

\subsection{Rule-Based Systems}
\label{sec:Rule Based Systems}

Cassell \etal\cite{Cassell_1994} presented one of the early studies on rule-based framework to synthesize behaviors. They defined several rules to generate appropriate behaviors dictated by the meaning of the message. In a later study, Cassell \etal \cite{Cassell_2004} introduced 
the \emph{behavior expression animation toolkit} (BEAT), which uses text to create animations with appropriate and synchronized gestures. BEAT tags semantic labels in the text, which are mapped into appropriate behaviors by heuristics rules suggested after observing human-human nonverbal displays. The synchronization is decided based on the timing of the words in the text. Poggi \etal \cite{Poggi_2005} introduced GRETA, which is an \emph{embodied conversational agent} (ECA), comprising several modules such as emotional mind, dialog manager, plan enrichment and body generator. The body enrichment module labels text with appropriate behaviors assigning synchronization points, which are realized by the body generator module.  GRETA includes a number of predefined gestures which can be exploited to generate animations with specific communicative goals. Kopp and  Wachsmuth \cite{Kopp_2004} proposed to find a prominent word or phrase to synchronize speech and gestures. The prominent words convey the communicative goal, creating anchors for the peak of the gesture. Marsella \etal \cite{Marsella_2013} proposed a framework to generate animation from speech. Their system uses an \emph{automatic speech recognition} (ASR) module to get the transcriptions, which are semantically analyzed to extract communicative goals in the message. They defined a list of behaviors associated with the communicative goals, mapping the text to behaviors. Their system also analyzes emotional cues in speech extracting arousal level, which dictates the selection of the behaviors generated for each communicative goal.

\subsection{Data-Driven Systems}
\label{sec:Data Driven Systems}

An alternative approach to generate behaviors is using data-driven methods that exploit the relationship between body movements and acoustic features (e.g., prosody). Vigot \etal \cite{Voigt_2014} demonstrated that there is a statistically significant correlation between prosodic features and raw body movements. Graf \etal \cite{Graf_2002} showed that there is correlation between prosodic events and behaviors such as eyebrow and head movements. Busso \etal \cite{Busso_2007_3} reported that the correlation between prosodic features and head movements across different emotions are on average more than $\rho=0.69$, using \emph{canonical correlation analysis} (CCA). Speech and gestures also co-occur. The study from McNeill \cite{McNeill_1992} showed that more than 90\% of the gestures occur while speaking, showing the tied connections between these modalities. These results have motivated synthesizing behaviors using speech-driven models.

Busso \etal \cite{Busso_2007} proposed emotion dependent \emph{hidden Markov models} (HMM) to synthesize head movements with prosodic features. Mariooryad \etal \cite{Mariooryad_2012_2} investigated several \emph{dynamic Bayesian networks} (DBNs) to jointly model head and eyebrow movements driven by speech, capturing the dependencies not only between speech and facial behaviors, but also between head and eyebrow motions. Some studies have argued that speech is correlated with the kinematics of the behaviors. Le \etal \cite{Le_2012} proposed to jointly model prosodic features and kinematic features of head motion using \emph{Gaussian Mixture Models} (GMM). Levine \etal \cite{Levine_2010} presented a system to synthesize body movement using \emph{hidden conditional random fields} (HCRFs), modeling the relationship between prosody and kinematic features of the joint rotations. The task was to predict kinematic parameters from speech. They use reinforcement learning to select behaviors in the database that match the inferred kinematics parameters. Bozkurt \etal \cite{Bozkurt_2013} designed a system for generating upper body beat gestures based on prosodic features. They clustered prosodic features into intonational phrases, and movements into gestural phrases. These units were jointly modeled using a \emph{hidden semi Markov model} (HSMM), which allowed asynchrony between the gestures and prosodic phrases by modeling the state duration of the hidden state. Chiu \etal \cite{Chiu_2011} proposed to use \emph{hierarchical factored conditional restricted Boltzmann machines} (HFCRBMs) which learns how to generate the joint poses for the next frame based on the previous frame conditioned on the prosodic features.

\subsection{Hybrid Approaches}
\label{ssec:hybrid}

Rule-based and data-driven methods have advantages and disadvantages. Rule-based methods do not capture the range of behaviors observed during human interaction, are limited by the stored behaviors, and often result in repetitive behaviors.  The synchronization between behaviors and speech is challenging, since they do not learn the synchronization from natural recordings. However, they can consider the meaning of the message to derive appropriate behaviors. Speech-driven methods can capture broader variations of behaviors, learning appropriate synchronization between speech and gestures. However, they may not create appropriate behaviors that match the intended communicative goal. Using pure speech-driven methods may be enough to predict beat gestures but not iconic or metaphoric gestures which are closely related to the message. Bridging the gap between rule-based and data-driven frameworks has the potential to create behaviors that are meaningful, timely synchronized, and representative of the range of variations observed during human interaction. 

Studies have attempted to combine both approaches creating hybrid frameworks. Stone \etal \cite{Stone_2004} designed a hybrid system to generate meaningful behaviors given the text. They jointly segment audio and motion capture recordings into units expressing pre-defined communicative intents. Given an input text, they parse the input into their predefined categories, using dynamic programming to find speech and motion capture segments that have the same communicative goal. The generation with this framework is limited to the stored speech segments. Sadoughi \etal \cite{Sadoughi_2014} proposed to constrain a speech-driven model based on the discourse functions of the sentence to generate more meaningful head and eyebrow motion. The study was limited to only two discourse functions: \emph{affirmation} and \emph{question}, where the subjective evaluation of the result showed improvements for the constrained model versus the unconstrained model when the constraint was \emph{question}.

\subsection{Contribution and Relation to our Prior Work}
\label{ssec:contributions}

This paper proposes a framework to create meaningful behaviors driven by speech. This framework creates animations based not only on prosodic features, but also on constraints which are either discourse functions (i.e., semantic structure of speech), or prototypical behaviors (e.g., head nods). This study builds upon our previous work, which we summarize in this section.

Sadoughi \etal \cite{Sadoughi_2014} proposed a model to generate speech-driven head and eyebrow movements constrained on discourse functions. The preliminary study tested the constrained model on one session of the IEMOCAP corpus, constraining the models on two dialog acts: \emph{question}, and \emph{affirmation}. The subjective evaluation of the models showed that the behaviors from the constrained models are more preferable, natural and appropriate compared to the unconstrained model. In Sadoughi and Busso \cite{Sadoughi_2015_2}, we explored the idea of constraining the models using prototypical behaviors such as head nods. The models were trained with gestures directly retrieved from the corpus by providing few examples. 

The models presented in our preliminary studies have several limitations. First, the variability of the generated behaviors is limited since the model optimization is susceptible to a poor initialization to reduce the mean square error, often resulting in average trajectories. Second, the structure of the proposed models requires balanced datasets per constraint, which is an unnecessary restriction. This study presents an improved speech-driven model that overcomes these problems by changing the structure of the model and the training strategy, which systematically reduces the confusion between the constraints during training. The contributions of this paper are (1) designing a constrained speech-driven model to generate more meaningful behaviors (Sec. \ref{ssec:cm1}), (2) an initialization technique which increases the range of motion associated with the behaviors (Sec. \ref{ssec:Initialization}), and (3) a novel training approach to effectively learn distinct patterns associated with different constraints (Sec. \ref{ssec:SparseMatrix}). 

\section{Overview}
\label{sec:overview}

This study aims to improve nonverbal displays of CAs using speech-driven models that are constrained by either the underlying discourse function in the message or prototypical behaviors specified by rule-based systems. In an attempt to create meaningful gestures, Marsella \etal \cite{Marsella_2013} defined several functions based on the content of the speech. These discourse related functions create a mapping between content and behaviors. Likewise, Poggi \etal \cite{Poggi_2005} designed a toolkit with several embedded functions to generate behaviors. The inputs to these mappings are communicative goal of the utterance, which we call discourse functions. These discourse functions are associated with relevant gestures that contribute in understanding the underlying message of the speech.

\begin{figure}[t]
\centering
 \includegraphics[trim = 8mm 97mm 8mm 14mm, clip, width=1.0\columnwidth]{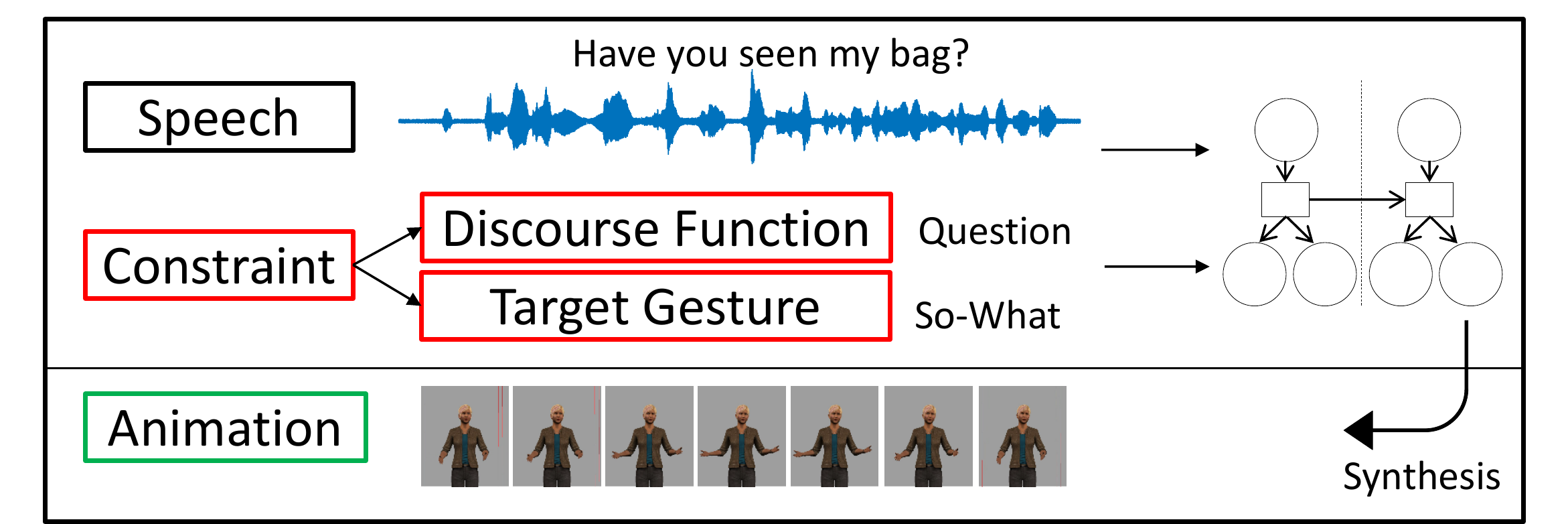}
\caption{Overview of the proposed system to generate behaviors. In addition to speech, the models take constraints as input (either discourse functions of target gesture), generating meaningful data-driven behaviors.}
\label{fig: flowchart}
\end{figure}

Figure \ref{fig: flowchart} gives the overview of our system, which takes as input speech and the underlying discourse function or intended gesture, producing meaningful behaviors that are timely synchronized with speech, convey the right message, and display the range of behaviors observed during human interactions. Our framework can take the role of behavior realizer proposed under the SAIBA framework \cite{Kopp_2006}, bridging the gap between rule-based and data-driven approaches. 

We aim to answer the following questions in the context of CAs,  ``do discourse functions affect behaviors?'' If so, ``is there a principled framework to capture the characteristics behaviors associated with discourse functions?'' Knowing the target gesture, ``can we effectively create the gesture which is synchronized with and modulated by speech?'' We address these questions by exploring four discourse functions: \emph{negation}, \emph{affirmation}, \emph{question} and \emph{suggestion}. The analysis in Section \ref{sec:Analysis} reveals that the behaviors observed during these discourse functions are in fact different. We propose a principled speech-driven approach to capture the characteristic behaviors for each discourse function. We also propose to constrain the models with prototypical behaviors. While the framework is general, we evaluate three hand gesture and two head gestures.

\section{Resources}
\label{sec:geswmean}

This section provides a brief description of the corpus used in this study, focusing on the annotation process and the method used to retrieve target gestures. 

\subsection{The MSP-AVATAR Database}
\label{sec:Design}

We collected the MSP-AVATAR corpus \cite{Sadoughi_2015} to study the role of discourse functions and gestures. This corpus was collected to provide data to synthesize more meaningful and naturalistic behaviors. 

The MSP-AVATAR corpus contains recordings of dyadic interactions based on improvisations of daily scenarios. It encompasses the recordings from six actors interacting in four dyadic interactions. The scenarios are carefully designed such that they include the use of eight discourse functions: \emph{contrast}, \emph{confirmation/negation}, \emph{question}, \emph{uncertainty}, \emph{suggestion}, \emph{giving orders}, \emph{warning}, and \emph{informing}. There are also scenarios prompting the actors to use iconic gestures (e.g., large, small) and deictic gestures for \emph{pronouns} (e.g., ``me'', ``you''). The discourse functions in this corpus are carefully chosen based on previous studies \cite{Marsella_2013,Poggi_2005}, which are likely to elicit characteristic behaviors.

\begin{figure}[t]
\centering
\subfigure[Upper-body markers]{
   \includegraphics[height=3.44cm]{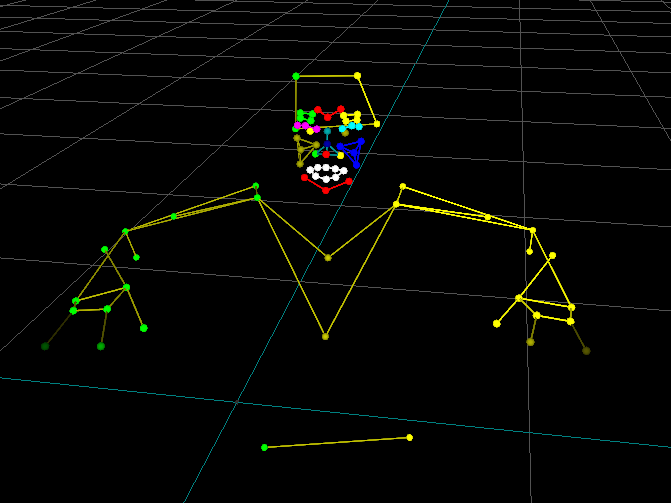}
   \label{fig: markers_a}
 }
 \subfigure[Actress with markers]{
   \includegraphics[height=3.44cm]{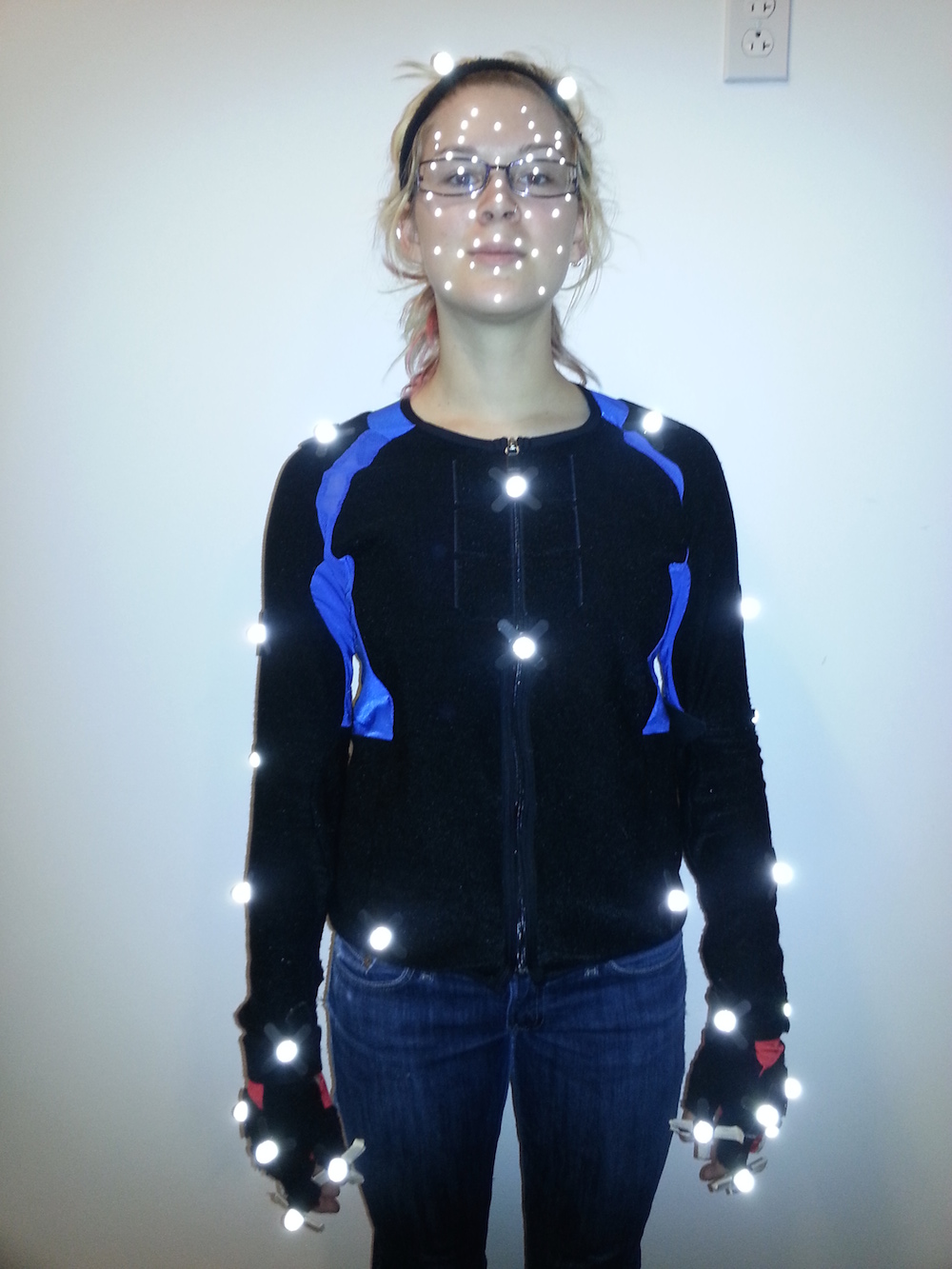}
   \label{fig: markers_b}
 }
\vspace{-0.03cm}
\caption{The MSP-AVATAR database \cite{Sadoughi_2015}. One of the subjects wore markers that were tracked with a VICON system.}
\label{fig: markers}
\end{figure}

The corpus consists of audio, video and motion capture recordings, collected at the Motion Capture laboratory of the University of Texas at Dallas. In each dyadic session, one of the actors wore 43 facial markers, and a suit in which we attached 28 markers (Fig. \ref{fig: markers}). Therefore, we have motion capture data for four different subjects. The facial markers include most of the \emph{feature points} (FPs) in the MPEG-4 standard \cite{Pandzic_2002}. For the upper body, we follow the position of the markers in the \emph{Vicon skeleton template} (VST). For each of the actors, we used a Lavalier microphone (SHURE MX150) connected to a portable digital recorder (TASCAM DR-100MKII). The microphone recorded the speech at a resolution of 16 bit and a sampling rate of 44.1 KHz. We used two Sony handycams HDR-XR100 which recorded at 1,920 $\times$ 1,080 resolution. The cameras were positioned to record the frontal view of each actor, without interfering with the Vicon system. In total, we have 74 sessions with a duration of two hours and fifty eight minutes. 

\subsection{Annotation of Discourse Functions}
\label{sec:annotation-of-discourse-function}
We manually annotated the 74 sessions, identifying sentences associated with discourse functions. Some of the discourse classes are harder to reliably annotate, so we only consider four classes: asking questions (\emph{question}), showing agreement (\emph{affirmation}), showing disagreement (\emph{negation}), and making suggestions (\emph{suggestion}). The evaluation was conducted with \emph{Amazon mechanical turk} (AMT), using the OCTAB interface designed by Park \etal \cite{Park_2012}. This toolkit is suitable for segmental annotations of the videos, where annotators can mark the beginning and end of segments in the videos where they noticed the requested discourse function. To improve the quality of the annotations, our approach identifies good evaluators using a screening phase.  We ask the evaluators to annotate the discourse function \emph{questions}, which is one of the easiest tasks. We also annotated this discourse function in our laboratory. We manually compared the annotations provided by each evaluator with our annotations, selecting the ones who provided reasonable answers. Then, we invited the selected evaluators to complete the rest of the assignments for the other three discourse functions. We recruited three annotators per assignment. 

We use the method proposed by Zhou \etal \cite{Zhou_2014} to aggregate the annotations coming from different annotators. This method solves a crowdsourcing model which estimates the hidden variables relevant to the difficulty of the tasks and the hidden variables relevant to the reliability of the annotators by using the minimax conditional entropy principle. The approach not only provides the hard labels after fusion, but also gives a confidence level in the assigned label (i.e., the soft assignment). To use this method, we consider our task as a binary classification where each video frame either belongs to the target discourse function or not (30fps). We derive a soft assignment for each of the four discourse functions using the three evaluations per frame.
We only consider the frames whose soft assignments are more than 0.9 for one of the discourse function, increasing the reliability in the labeled segments. Notice that the annotated frames are not mutually exclusive among the discourse functions. If we separately consider the co-occurrences between two or more discourse functions (e.g. \emph{suggest} and \emph{question}) as extra constraints, we would need enough realizations of these combinations. Unfortunately, the total durations of the co-occurrences of labels between 2 or more discourse functions vary between 0.5s to 310s, which is not enough. For simplicity, we enforce mutually exclusive segments by removing the overlaps, keeping as many segments as possible. This approach results in total durations of 734.4s for \emph{affirmation}, 1,118.7s for \emph{negation}, 1,149.1s for \emph{question}, 1,582.5s for \emph{suggestion}, and 6,111.7s for \emph{other}.

\begin{figure*}
\centering
\subfigure[So-What]
{
  \includegraphics[height=1.1in]{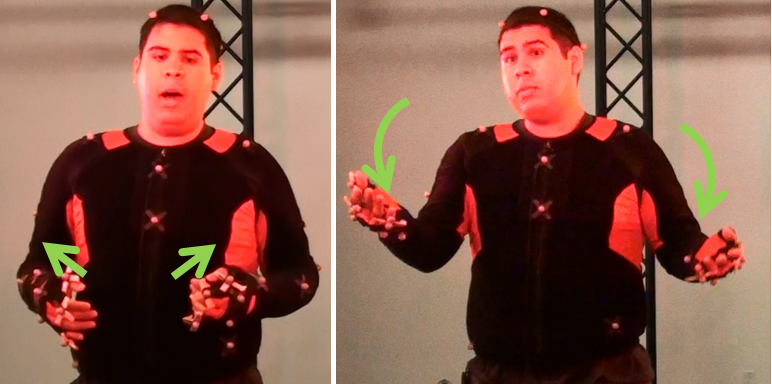}
  \label{fig:Gesture_a}
}
\subfigure[To-Fro]
{
  \includegraphics[clip, height=1.1in]{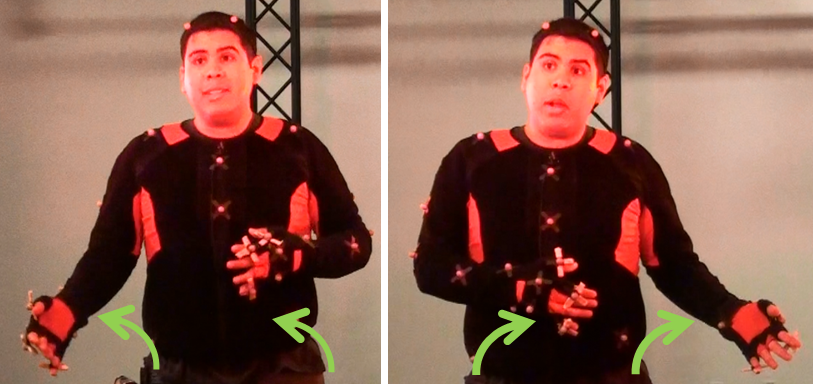}
  \label{fig:Gesture_b}
}
\subfigure[Regress]
{
  \includegraphics[clip, height=1.1in]{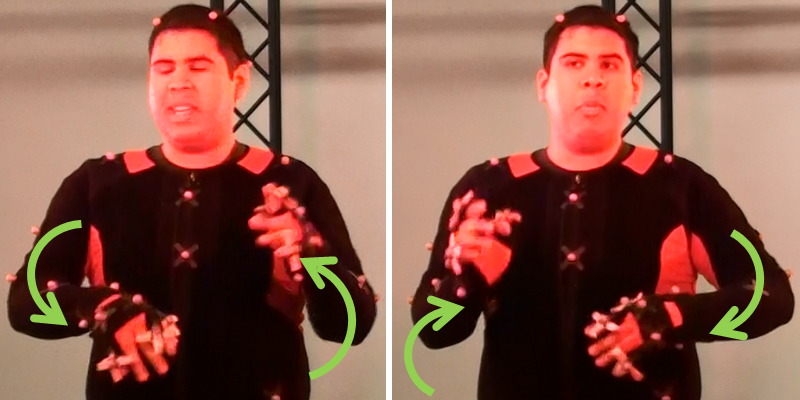}
  \label{fig:Gesture_c}
}
\vspace{-0.3cm}
\caption{Illustration of the three prototypical hand gestures considered in this study. These gestures were defined by Kipp \cite{Kipp_2003}.}
\vspace{-0.2cm}
\label{fig: HandGestures}
\end{figure*}


\subsection{Motion \& Audio Features}
\label{ssec:features}
The data-driven models take speech features as input, generating the most likely behaviors. This study considers head and hand gestures. We use the upper body joint rotations derived after solving the skeleton of the motion capture recordings in Blade. We consider the pitch, yaw, and roll rotations for the head (i.e., 3 \emph{degree of freedom} (DOF)),  arms (3 DOF $\times$ 2) and forearms (2 DOF $\times$ 2), normalizing these features using z-normalization per subject. The sampling rate for the motion capture data is 120fps.

The acoustic features correspond to prosodic features, following our previous work \cite{Busso_2007,Mariooryad_2012_2, Sadoughi_2014,Sadoughi_2016}. We extract the fundamental frequency and energy using Praat \cite{Boersma_1996}, estimating their first and second order derivatives resulting in a 6D feature vector. These features are extracted using 40ms windows every 16.67ms with 23.3ms overlap (i.e., 60fps). We interpolate the unvoiced segments in the fundamental frequency to avoid discontinuities. The feature vector is up-sampled to match the sampling rate of the motion capture data (i.e., 120fps).

\subsection{Prototypical Behaviors}
\label{ssec:Prototypical}

This study demonstrates that it is possible to create prototypical behaviors using data-driven models. While the framework is general, we only consider three prototypical behaviors for hand and two prototypical behaviors for head movements. Figure \ref{fig: HandGestures} illustrates the target hand gestures. The behaviors are defined as follows:

\noindent
\underline{\emph{Head Nod:}} One or more pitch rotations of head.

\noindent
\underline{\emph{Head Shake:}} One or more yaw rotations of head.

\noindent
\underline{\emph{To-Fro:}} Movement of both hands form side to side.

\noindent
\underline{\emph{So-What:}} Movement of both hands in an arc in an outward manner.

\noindent
\underline{\emph{Regress:}} Movements of hands in circles towards the body.

\begin{table}
\centering
\caption{Prototypical gesture considered in the behavior retrieval framework. Number of examples in the train, and test \& develop sets.}
\begin{tabular}{c c| c| c}
\hline
 Region & Behavior &\#$\mathit{Samples}_\mathit{Train}$ & \#$\mathit{Samples}_\mathit{Test\&Dev}$\\
\hline
\hline
\multirow{2}{*}{HEAD} 
& Nod  & 56 & 308\\
& Shake & 39 & 237\\
\hline
\multirow{3}{*}{HAND} 
& To-Fro & 47 & 77\\
& So-What & 28 & 72\\
& Regress & 24 & 114\\
\hline
\end{tabular}
\label{tab:behs}
\end{table}

The data-driven models require enough examples of these gestures to effectively train the models. We use the supervised framework introduced by Sadoughi and Busso\cite{Sadoughi_2015_2} to automatically retrieve these instances from the dataset. The key idea is to annotate few examples of the target behavior, and retrieve the rest of the segments until we have enough data to train the models. The approach is a supervised approach that simultaneously solves the segmentation and detection of the target gestures. The first step is downsampling the motion capture sequences using clusters. This is a nonuniform downsampling approach that discards segments without variations while keeping changes in the trajectories. Then, we use a multi-scale sliding window framework that considers windows of different sizes, accounting for variation in the duration of the gestures. The next step is to determine whether the selected segments include the target gesture. The approach consists of two steps. In the first step, we screen the segments using one-class \emph{support vector machine} (SVM), which reduces the potential segments, removing everything that departs from trajectories of the training examples. The second step uses the \emph{dynamic time alignment kernel} (DTAK) algorithm to evaluate the candidate segments in more detail. For DTAK, we use the implementation provided by Zhou \etal\cite{Zhou_2008}.

In Sadoughi and Busso\cite{Sadoughi_2015_2}, we set the detection threshold by maximizing the f-score. However, for this study it is more important that the selected segments are indeed from the target gestures (i.e, recall rate is less important). Therefore, in this study we set the detection thresholds per subject by maximizing the precision on the developing set.

\begin{table}[t]
\caption{The precision rates of the retrieved gestures on the test set.}
\centering
\begin{tabular}{c c| c c} \\
\hline
Region & Behavior &\multicolumn{2}{c}{$\mathit{Samples}_\mathit{Test\&Dev}$} \\ 
\cline{3-4}
 && Precision & Retrieved gestures  \\
 &&[\%]&[\#]\\

\hline
\hline
\multirow{2}{*}{\rotatebox[origin=c]{90}{Head}} 
& Shake 		& 97.10 & 69\\ 
& Nod 		& 96.55 & 87 \\ 
\hline
\multirow{3}{*}{\rotatebox[origin=c]{90}{Hands}} 
& So-What 	 & 80.00 & 20\\ 
& To-Fro 		& 68.18 & 22\\
& Regress  & 90.48 & 42\\\hline
\end{tabular}
\label{tab:res_det}
\end{table}

We manually annotated three sessions per subject to evaluate the behavior retrieval framework ($3\times4=12$).  Table \ref{tab:behs} gives the number of examples annotated per target behavior in these 12 sessions (column \#$\mathit{Samples}_\mathit{Test\&Dev}$). These 12 sessions are partitioned into development (two session per speaker) and test (one session per speaker) sets using three-fold cross-validation. The development set is exclusively used to set the detection thresholds. Table \ref{tab:res_det} shows the accuracy of the behavior retrieval framework. The precision rates for head gestures are higher than 96\%. For hand gestures the precision rates are higher for \emph{so-what} (80\%) and \emph{regress} (90.5\%). The precision rate is lower for the \emph{to-fro}, since the behaviors are more complex. 

Since our algorithm independently solves the detection of gestures, it is possible to have overlaps between two or more target gestures. We observe that the durations of these overlaps are 552.4s for head gestures, and between 30.2s to 91.3s for hand gestures. Similar to the annotation of discourse functions, we separately remove the overlap segments, resulting in mutually exclusive segments for hand and head gestures. For head gestures, we identify 1029.9s for \emph{shake}, 2056.3s for \emph{nod}. The remaining frames are labeled as \emph{other} (7484.1s). For hand gestures, we identify  201.6s for \emph{so-what}, 448.6s for \emph{to-fro}, and 567.3s for \emph{regress}. The remaining frames are labeled as \emph{other} (9352.7s).

\section{Statistical Analysis of the Constraints}
\label{sec:Analysis}

Before constraining our models on either discourse functions or target behaviors, we explore whether the hand and head behaviors vary across different categories of the constraints (e.g., differences in head motion for \emph{questions} and \emph{affirmations}). If all the different discourse functions or target behaviors do not have any effect on the behaviors, there is no value in constraining our speech-driven model, so this analysis is relevant. 

We use statistical tests to evaluate whether the presence of different discourse functions creates a significant difference in the behaviors. We consider pitch, yaw, and roll rotations for the head and joint rotations for the arms and forearms. The Kolmogorov-Smirnov test indicates that the distribution of the data is not normal, so we rely on the \emph{Kruskal-Wallis} (KW) test for the evaluation, asserting significance at $p$-value$<$0.05. When the KW test rejects the null hypothesis, meaning that at least two of the distributions are not the same, we use \emph{Dunn \& Sid\'{a}k's approach} (DSA) to perform pairwise comparisons to identify which pairs are different. Table \ref{fig:stat} gives the results for the pairs whose distributions are statistically different. The behaviors associated with the four discourse functions are different across different regions. For instance, \emph{negation} and \emph{affirmation} show differences in head yaw and pitch rotations, but not in roll rotation. Also, there are differences between arm rotations for both hands when the sentence is either an \emph{affirmation} or a \emph{suggestion}.

\begin{figure}

	\includegraphics[trim = 0cm 0cm 0cm 0cm, clip,width=4.2cm]{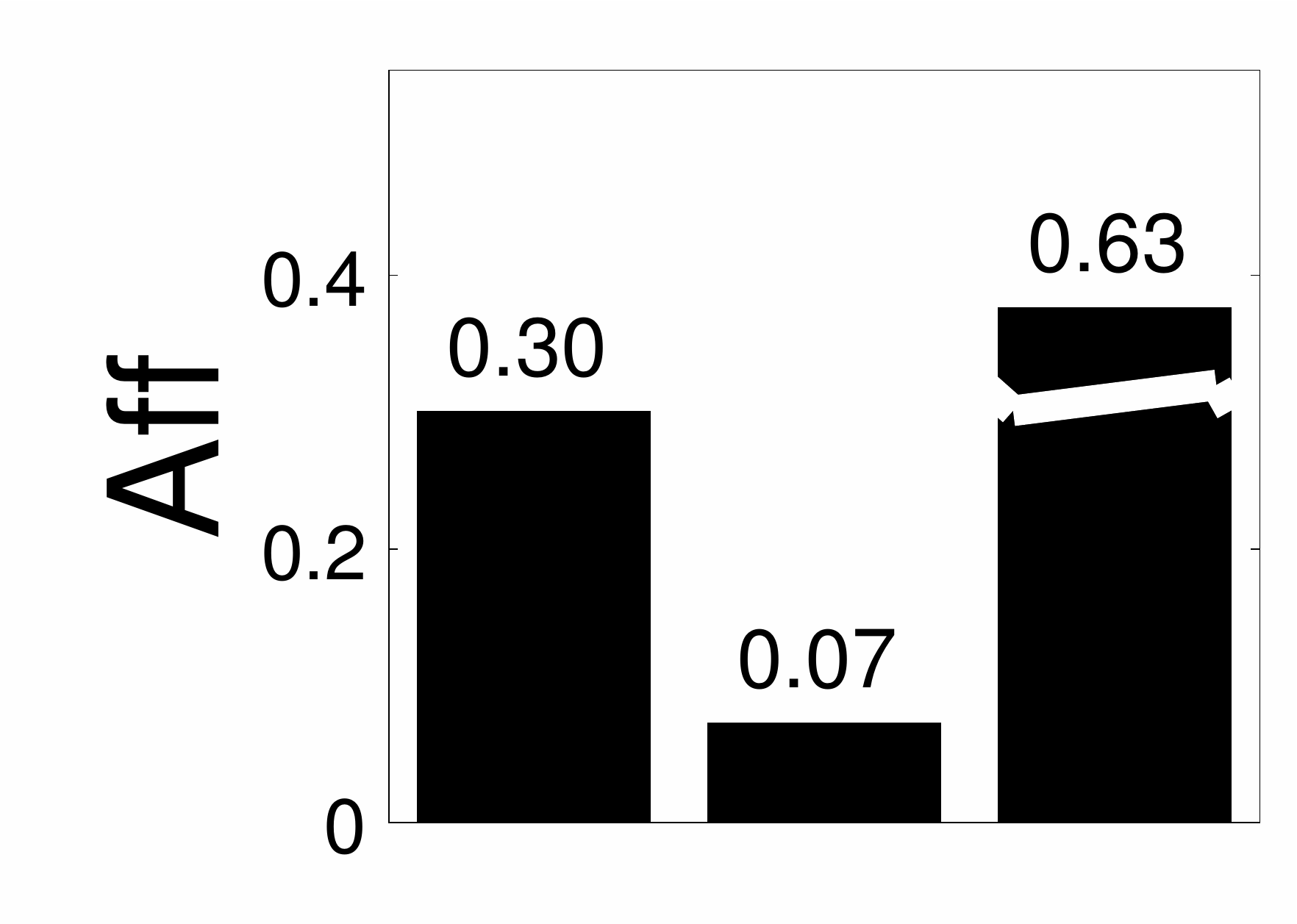} \includegraphics[trim = 3.1cm 0cm 0cm 0cm, clip,width=4.6cm]{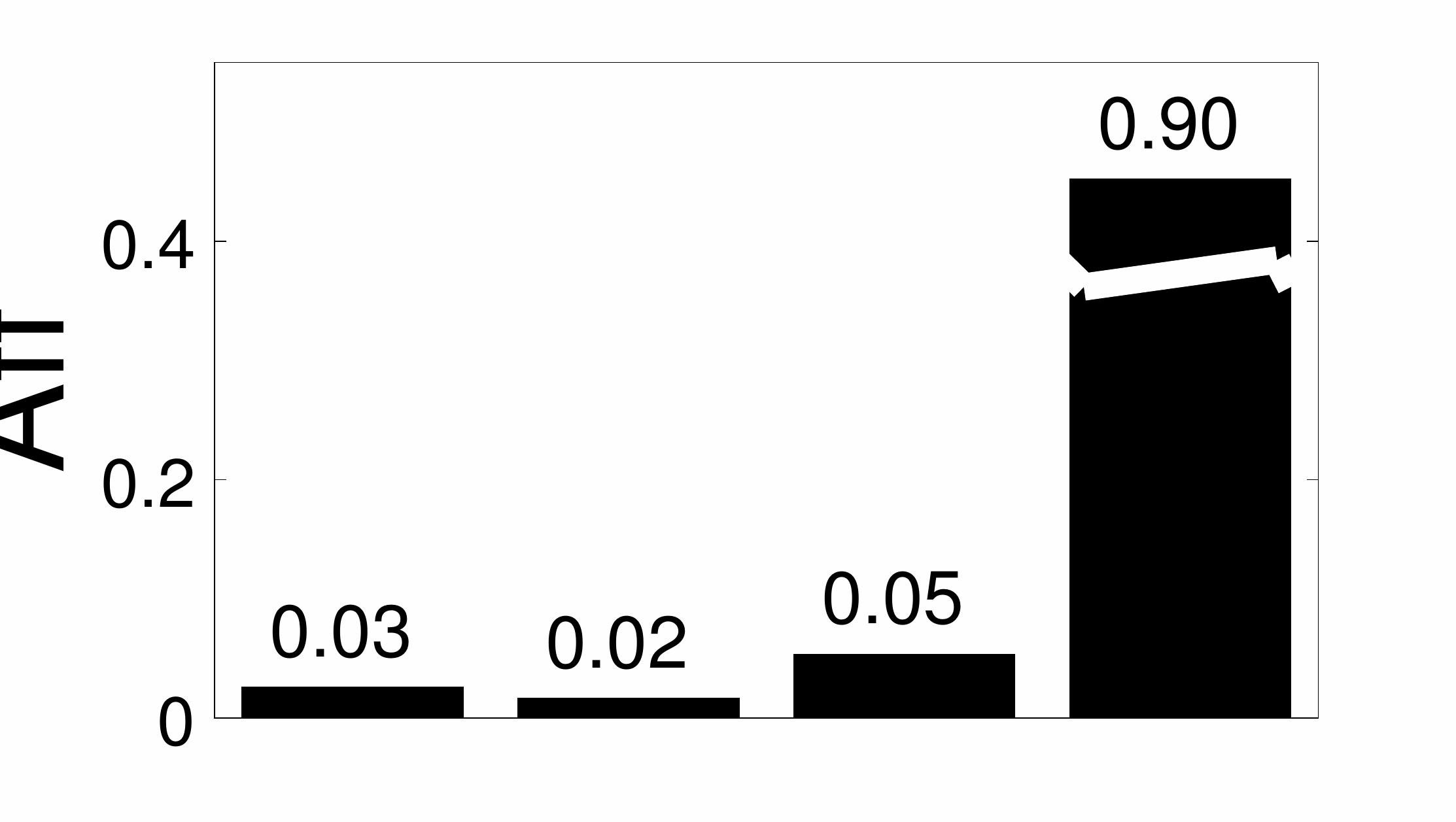}
	\includegraphics[trim = 0cm 0cm 0cm 0cm, clip,width=4.2cm]{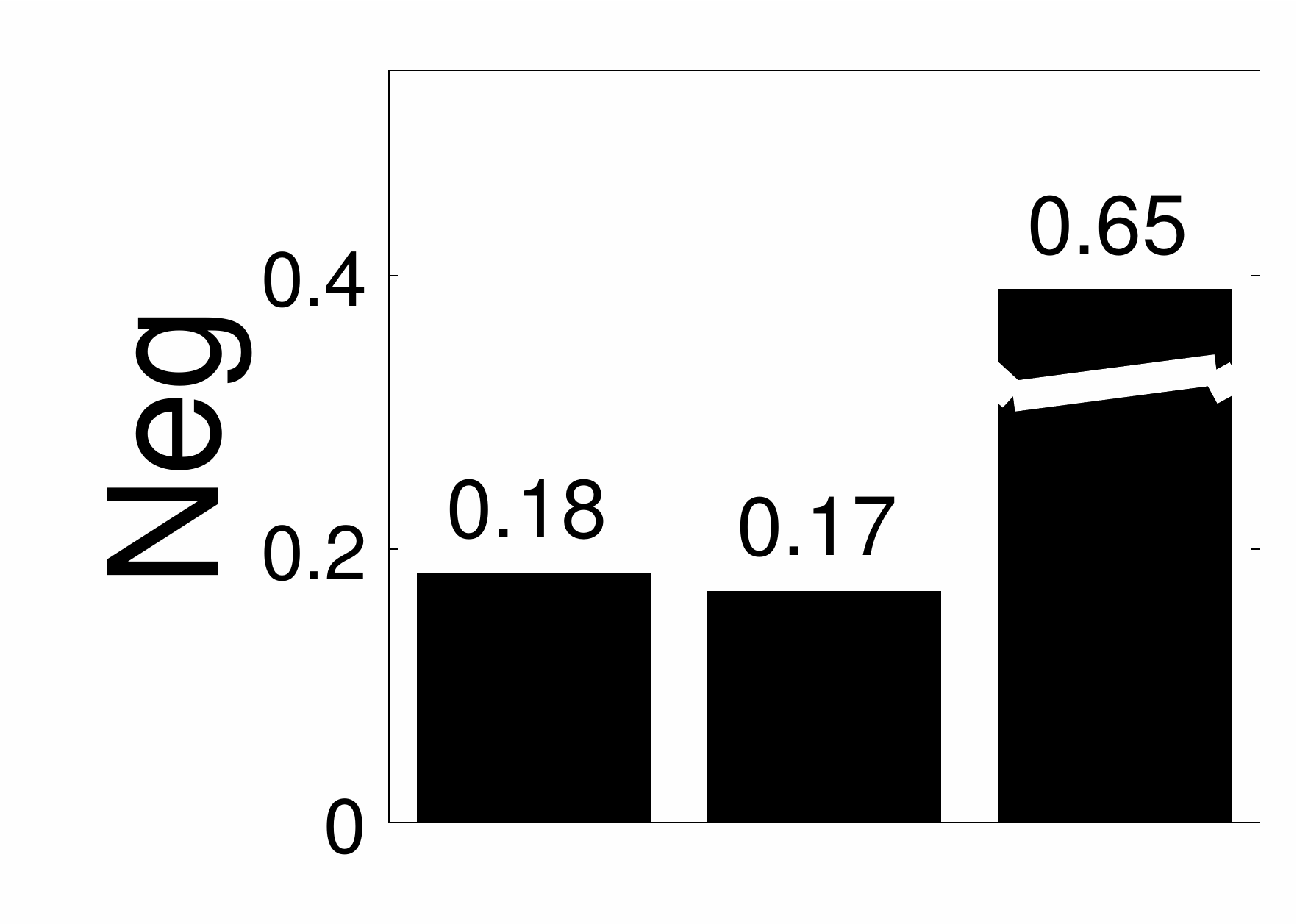} \includegraphics[trim = 3.1cm 0cm 0cm 0cm, clip,width=4.6cm]{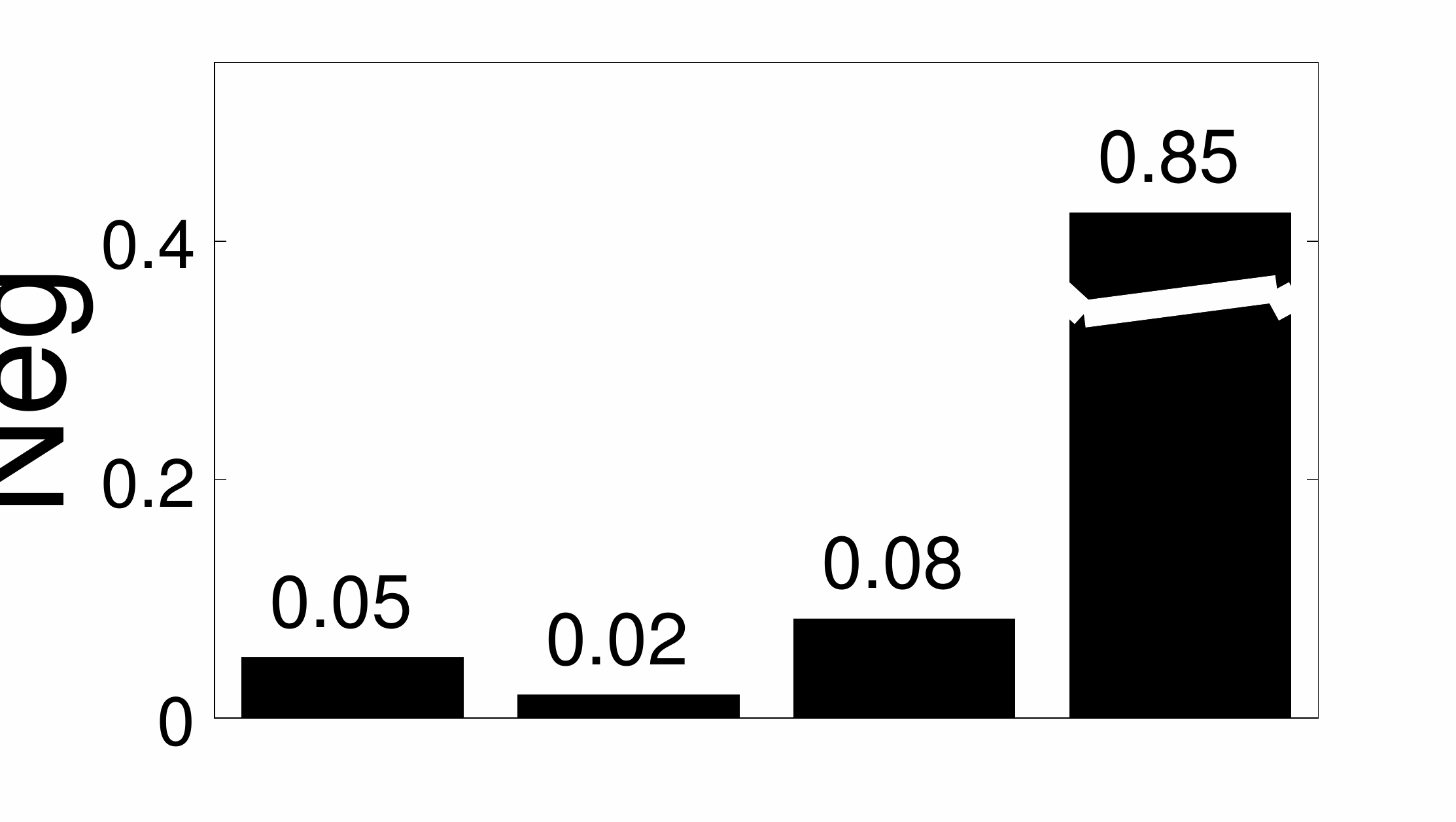}
	\includegraphics[trim = 0cm 0cm 0cm 0cm, clip,width=4.2cm]{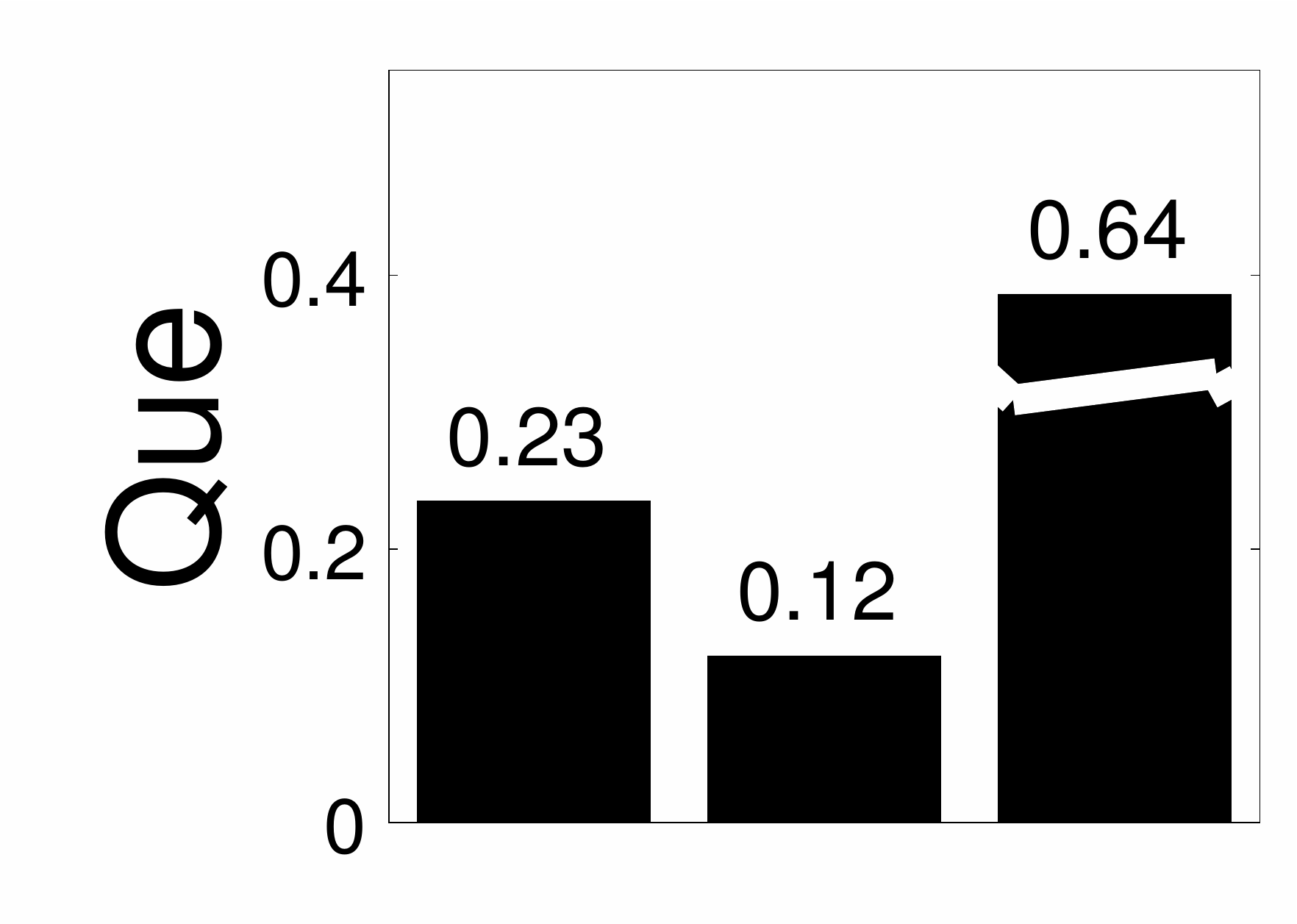} \includegraphics[trim = 3.1cm 0cm 0cm 0cm, clip,width=4.6cm]{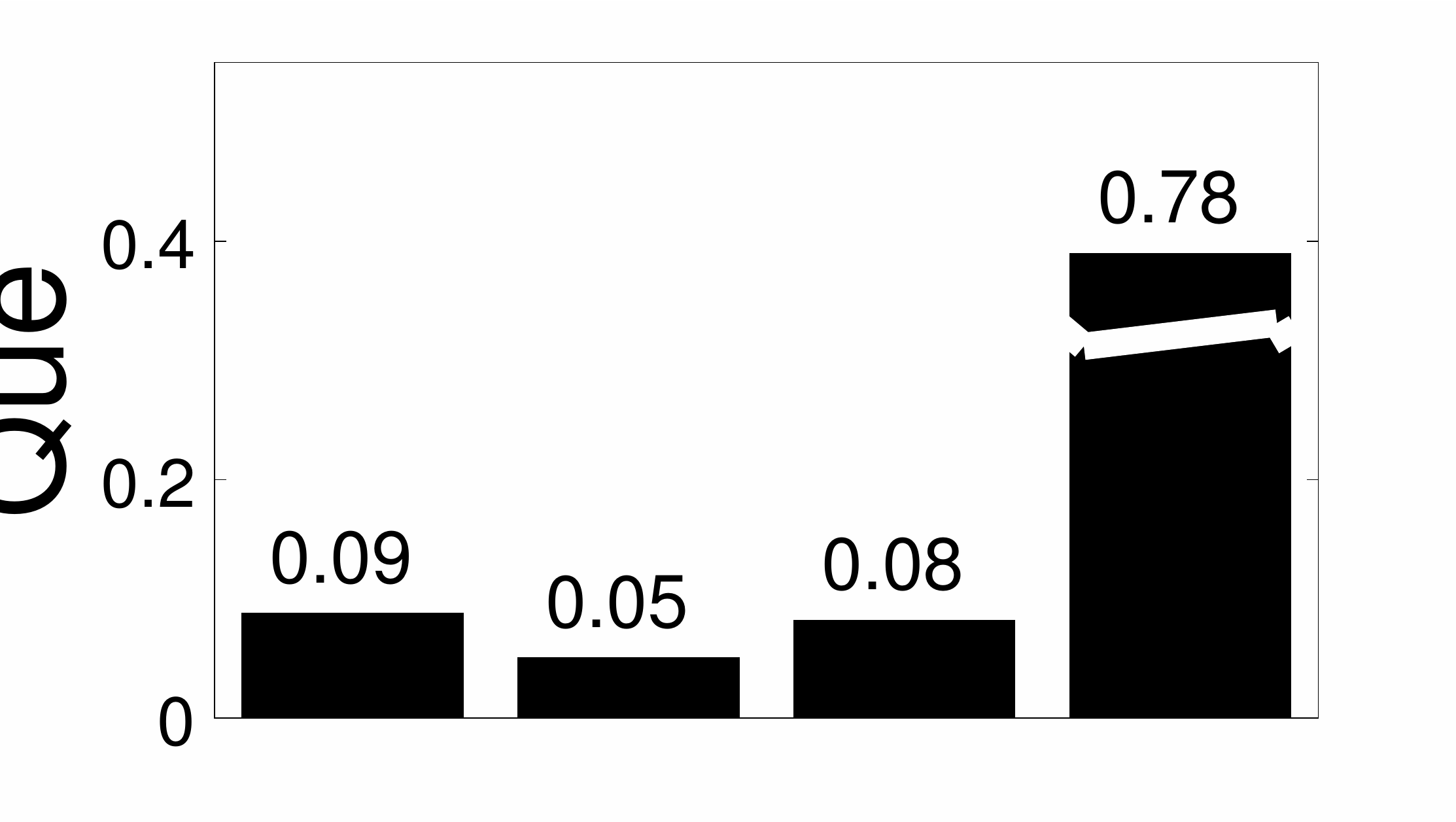}
	\includegraphics[trim = 0cm 0cm 0cm 0cm, clip,width=4.2cm]{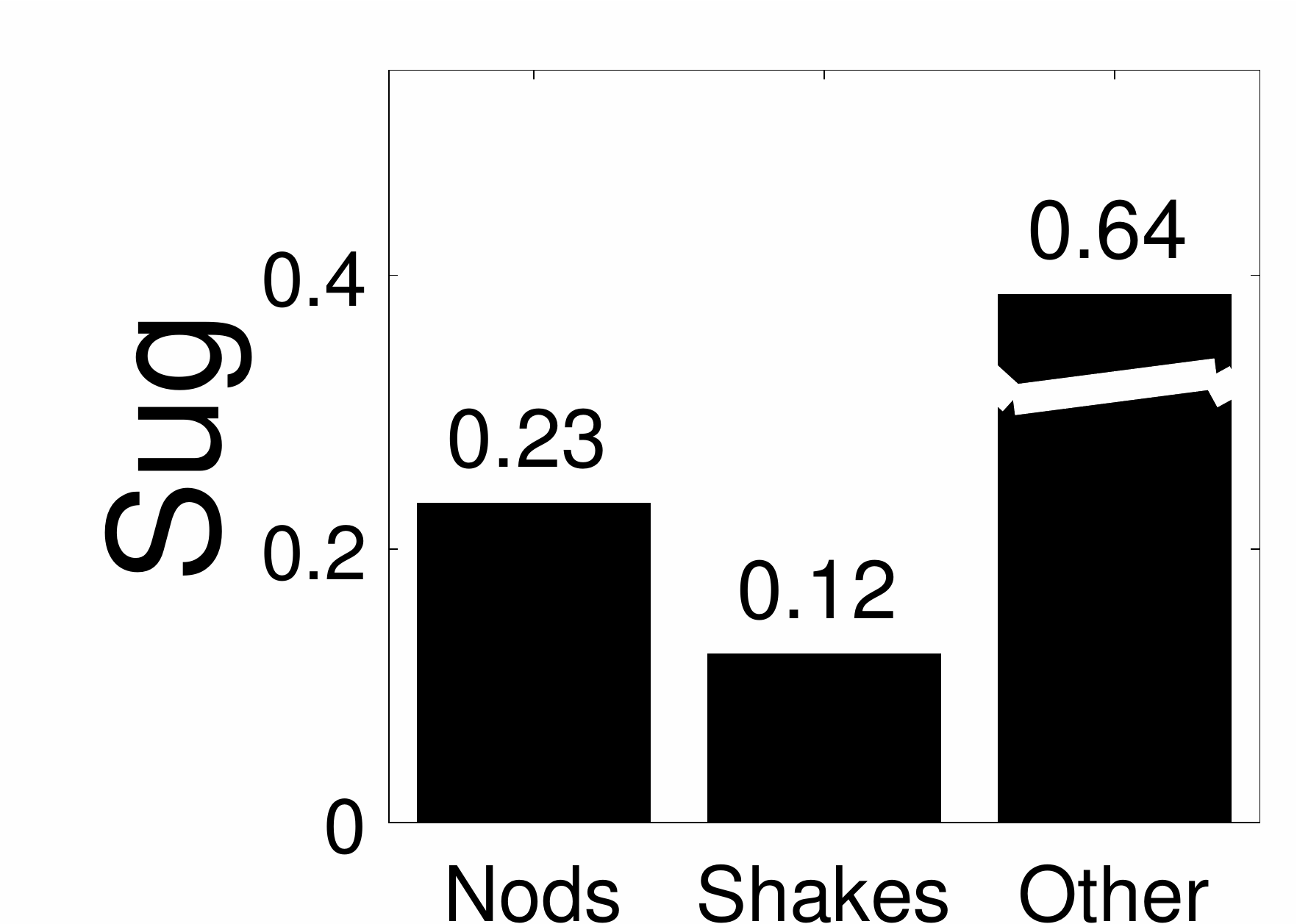} \ \ \ \includegraphics[trim = 4.1cm 0cm 0cm 0cm, clip,width=4.6cm]{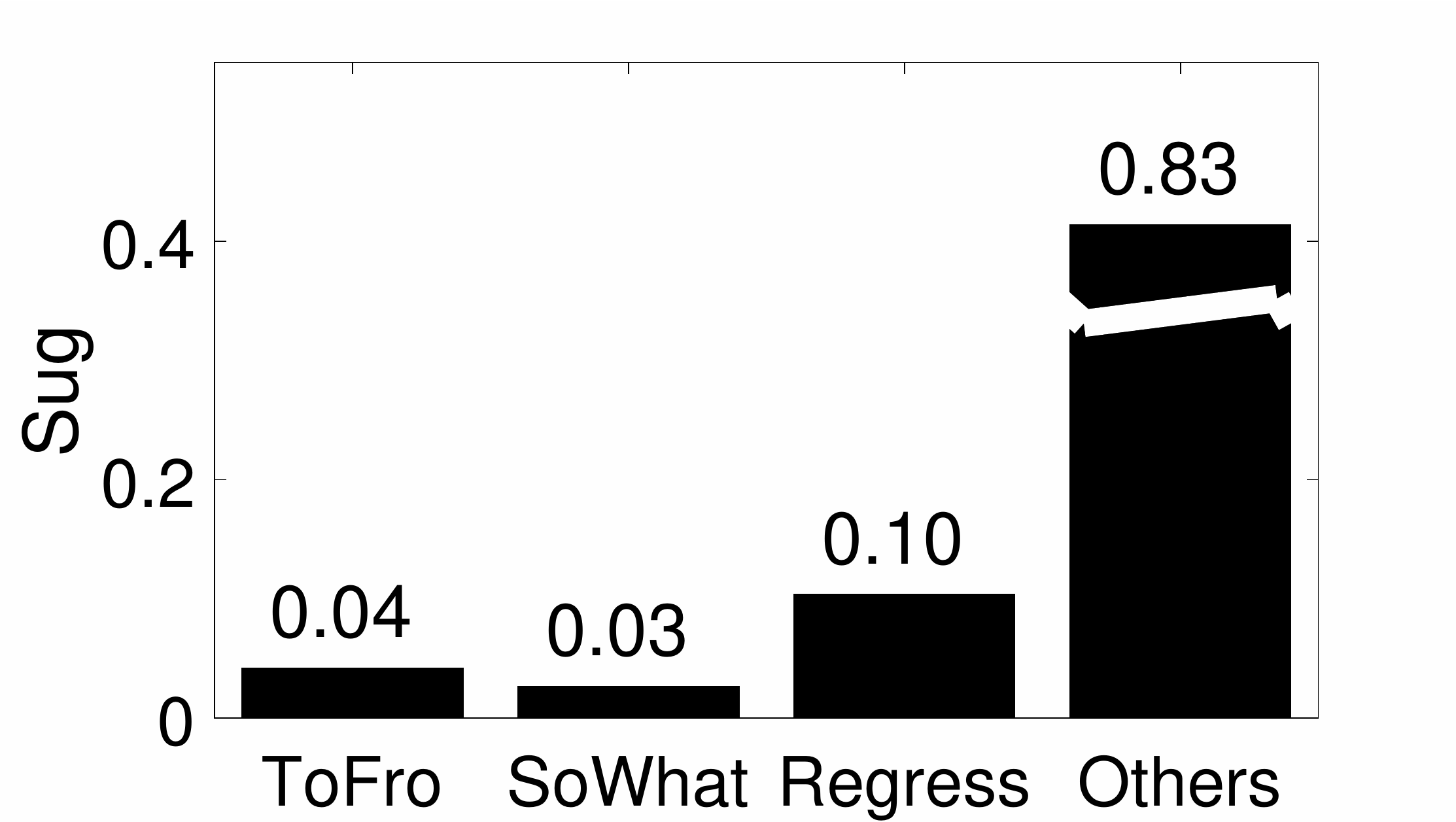}
\label{fig:hist}
\caption{The histograms of the behaviors for hand (left) and head (right) for each discourse function 
(\emph{Neg}: negations, \emph{Aff}: Affirmations, \emph{Que}: questions, \emph{Sug}: Suggestions, \emph{Other}: Other behaviors in the hand/head region).}
\label{fig:M1}
\end{figure}

We also analyze how discourse functions affect the prototypical behaviors for head and hand gestures. This analysis considers the frames assigned to each discourse function for each of the five prototypical gestures. Figure \ref{fig:hist} displays the normalized distribution of the target behaviors per discourse function, which is estimated by normalizing by the number of frames assigned to each discourse function (the addition of the distribution is 1 in each subfigure). This normalization is necessary, since some discourse functions are more frequent than others.  The figure displays separate results for head and hand movements. These histograms reveal some interesting relationships between discourse functions and behaviors. For example, the proportion of \emph{head shake} increases for \emph{negation}, whereas the proportion of \emph{head nod} increases for \emph{affirmation}. Some intuitive results are that the prototypical behavior \emph{so-what} occurs more often during \emph{questions}, \emph{regress} occurs more often during \emph{suggestions}, and \emph{to-fro} occurs more often during \emph{questions}. These histograms demonstrate that there are differences between the behaviors associated with discourse functions, which our models aim to capture. 


\begin{table}
\centering
\caption{Joint rotations that have significant differences using the KW test  asserting significance at $p$-value$<$0.05 (\emph{Neg}: negations, \emph{Aff}: Affirmations, \emph{Que}: questions, \emph{Sug}: Suggestions, \emph{Other}: None of the four discourse functions, \emph{r}: right, and \emph{l}:left).}
\begin{tabular}{l l l l}\\
\includegraphics[width=0.08\textwidth]{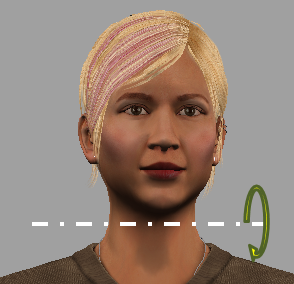}&\includegraphics[width=0.08\textwidth]{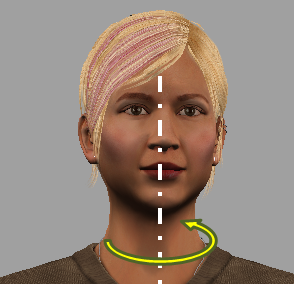}&\includegraphics[width=0.08\textwidth]{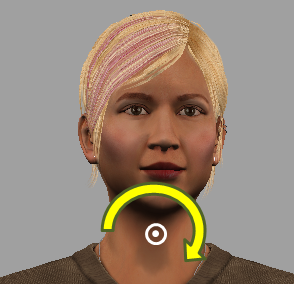}
&\includegraphics[trim = 10mm 40mm 70mm 70mm, clip, width=0.075\textwidth]{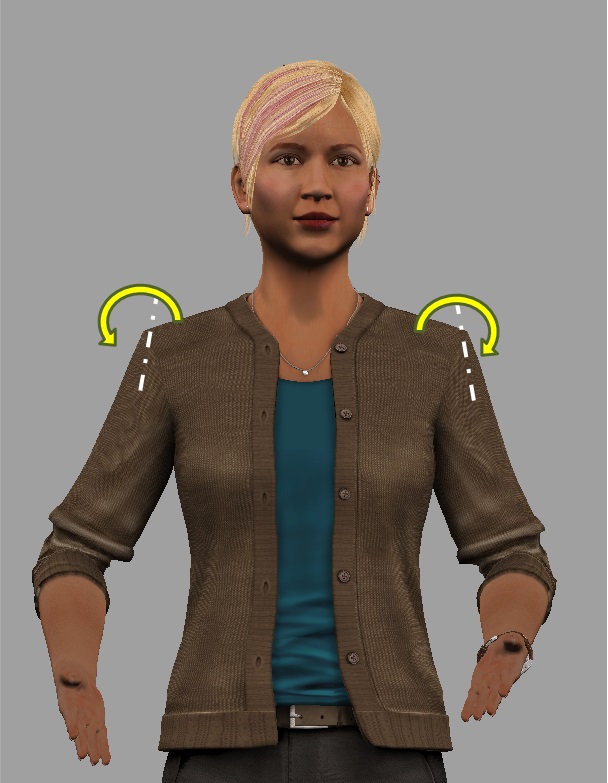}\\
\hline\hline
Neg-Aff & Neg-Aff & Neg-Sug& Sug-Other(r,l) \\
Neg-Que & Neg-Que & Aff-Sug& Neg-Que(r,l)  \\
Neg-Other & Neg-Sug & Aff-Other& Neg-Other(r,l) \\
Aff-Que & Neg-Other & Que-Sug& Aff-Que(r,l) \\
Aff-Sug & Aff-Que & Que-Other& Aff-Sug(r,l)  \\
Aff-Other & Aff-Other &         & Que-Other(r,l)\\
Que-Sug & Que-Sug &         & Neg-Aff(r)     \\
Que-Other & Que-Other &         & Aff-Other(l)	\\
Sug-Other & Sug-Other &         & Que-Sug(l)  	\\
\end{tabular}

\begin{tabular}{l l l l}\\
\includegraphics[trim = 10mm 40mm 70mm 70mm, clip, width=0.075\textwidth]{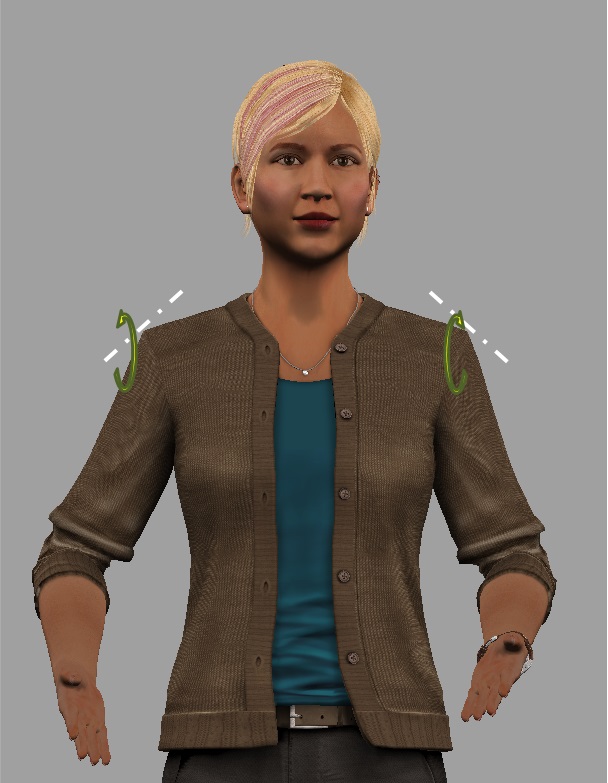}&
\includegraphics[trim = 10mm 40mm 70mm 70mm, clip, width=0.075\textwidth]{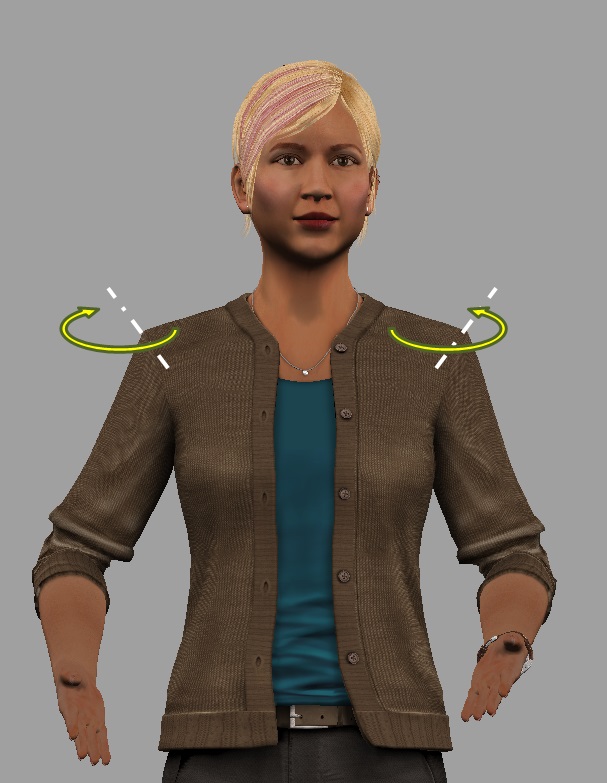}&
\includegraphics[trim = 0mm 5mm 93mm 120mm, clip, width=0.075\textwidth]{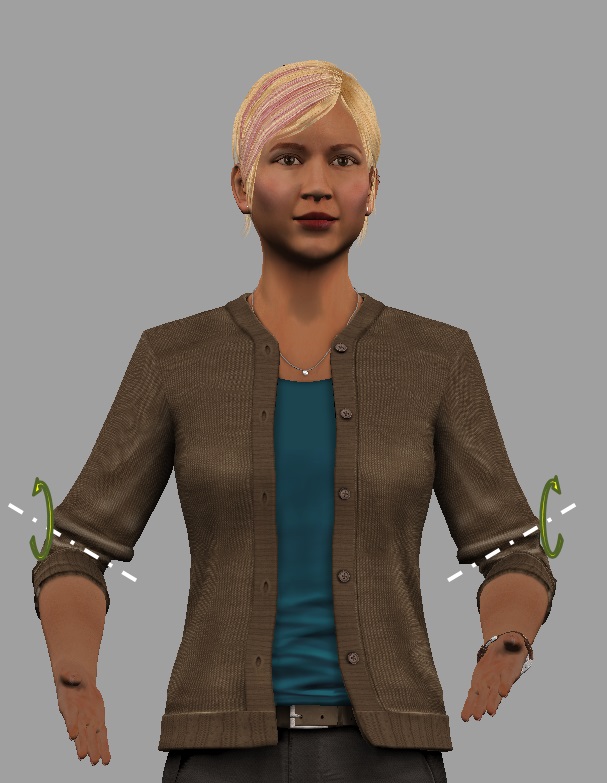}&
\includegraphics[trim = 0mm 5mm 93mm 120mm, clip, width=0.075\textwidth]{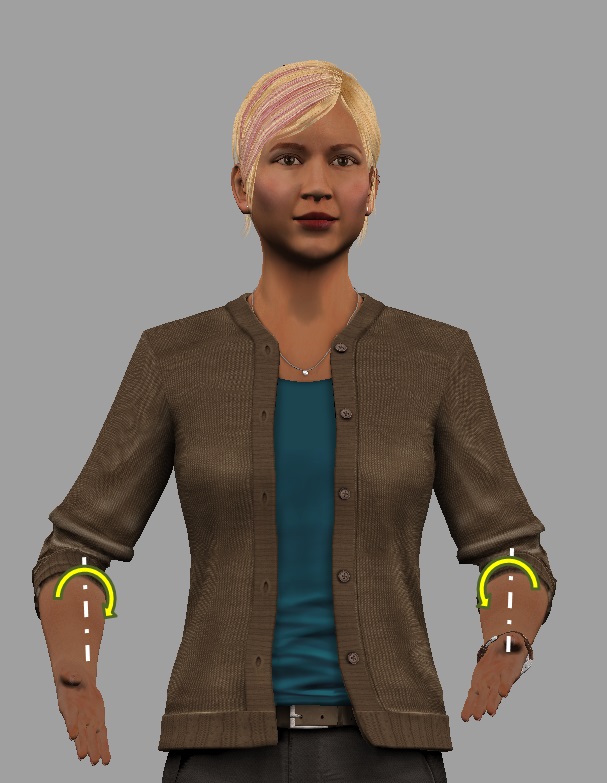}\\
\hline\hline
Neg-Aff(r,l) & Que-Sug(r,l) & Neg-Aff(r,l) & Neg-Aff(r,l) \\
Neg-Other(r,l) & Sug-Other(r,l) & Neg-Que(r,l) & Que-Other(r,l)\\
Aff-Que(r,l) & Aff-Sug(r,l) & Sug-Other(r,l) & Neg-Other(r,l) \\
Aff-Sug(r,l) & Aff-Other(r,l) & Neg-Other(r,l) & Aff-Que(r,l) \\
Aff-Other(r,l) &  Neg-Aff(r) & Aff-Que(r,l) & Sug-Other(r,l) \\
Que-Other(r,l) & Neg-Sug(r)  & Aff-Sug(r,l) & Que-Sug(r,l) \\
Sug-Other(r,l) & Neg-Que(l)  &Aff-Other(r,l) &  Neg-Que(r)\\
Neg-Que (l)  & Neg-Other(l)& Que-Other(r,l) &  Aff-Sug(r)\\
Neg-Sug (l)  & Aff-Que(l)  & Neg-Sug(r)     & Aff-Other(l)\\
  							&        	  &  Que-Sug(l)      &         \\
\end{tabular}
\label{fig:stat}
\end{table}

\section{Baseline Model}
\label{sec:Baseline}

We consider speech-driven methods built with \emph{dynamic Bayesian network} (DBN). This section introduces the original DBN proposed by Mariooryad and Busso \cite{Mariooryad_2012_2}, which is the building block of the proposed models. This DBN framework also serves as a baseline for our models.

\begin{figure}
\centering
\includegraphics[trim = 40mm 65mm 75mm 50mm, clip, width=6cm]{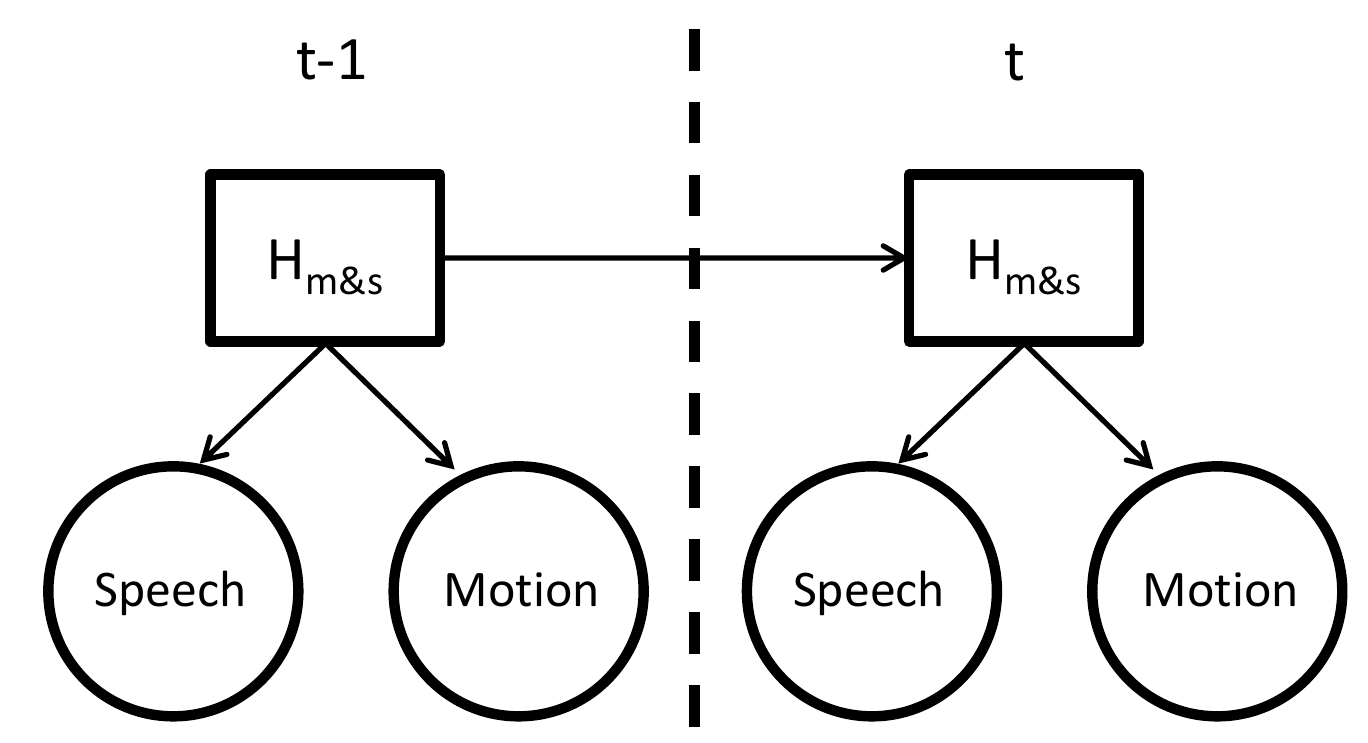}
\caption{Baseline DBN \cite{Mariooryad_2012_2}, which jointly models speech features and body movements (head or hand gestures in this paper).}
\label{fig:jDBN3}
\end{figure}

Figure \ref{fig:jDBN3} illustrates the baseline model, which was referred to as jDBN3 in Mariooryad and Busso \cite{Mariooryad_2012_2}. This structure was the best model to jointly capture not only the relation between speech and facial features, but also the relation between facial features. In the diagram, the circle nodes represent the observation variables and the rectangle nodes represent the hidden variables. In our model, the node \emph{Speech} represents the prosodic features and the node \emph{Motion} represents either hand or head motions. The nodes \emph{Speech} and \emph{Motion} are continuous variables and are modeled with Gaussian distributions. The hidden discrete state variable $H_{m\&s}$ represents the state configuration between speech features and the gesture. It serves as a discrete codebook constraining the speech and gesture space. The transition matrix between the hidden variables is ergodic, where the transition probabilities follow the Markov property of order one. The time unit of the DBN is the time frame in the data (120fps). 

  In this model, the nodes \emph{Speech} and \emph{Motion} are conditionally independent given $H_{m\&s}$. When the speech features are entered in the models, Bayesian inference updates the marginal probabilities of the state configuration node $H_{m\&s}$, affecting the node \emph{Motion}. This model preserves the full dependencies of the features within a modality by having full covariance matrices. This section describes the inference and synthesis method, emphasizing the improvement presented in this study, which enhances the range of movements synthesized by the model. 

\subsection{Inference}
\label{ssec:inference}

There are differences in the inference process for learning and synthesizing the gestures. During learning, we have access to the observations for the nodes \emph{Motion} and \emph{Speech}, so we use the full observation probability ($O^F_t(i)$) in Equation \ref{eq:fullO}. During synthesis, we only have observations for the variable \emph{Speech}, and the task is to predict the variable \emph{Motion}. Therefore, we use partial observation probability ($O^P_t(i)$) in Equation \ref{eq:partialO}. 

\begin{equation}
\begin{aligned}
O^F_t(i)=&P(Speech_t|H_{{m\&s}_t}=i) \\
\cdot &P(Motion_t|H_{{m\&s}_t}=i) \\
\end{aligned}
\label{eq:fullO}
\end{equation}

\begin{equation}
O^P_t(i)=P(Speech_t|H_{{m\&s}_t}=i)
\label{eq:partialO}
\end{equation}

\subsection{Synthesis}
\label{ssec:Synthesis}

During synthesis of the \emph{Motion} variable, we find the probabilities of the states $\gamma(i)$ at each time given the partial observation sequence and the model (Eq. \ref{eq:gamma}), using the Viterbi algorithm ($q_t$ is the state at time $t$, $O$ is the partial observation and $\lambda$ represents the parameters of the model). Equation \ref{eq:expectation} calculates the expected value for the node \emph{Motion} given speech features, where $\mu_{h}^i$ is the mean of the $i^{th}$ state for the variable \emph{Motion}.

\begin{equation}
\gamma_t(i)=P\left(q_t = i|O,\lambda\right)
\label{eq:gamma}
\end{equation}

\begin{equation}
\begin{aligned}
E\left[Motion_t|Speech\right] =
\sum_{i=1}^n\mu_{h}^i\gamma_t(i)
\end{aligned}
\label{eq:expectation}
\end{equation}

\subsection{Initialization of the States using Vector Quantization}
\label{ssec:Initialization}

\begin{figure}
\centering
       \subfigure[Random Initilaization]{
                \includegraphics[trim = 1cm 0cm 0.27cm 0cm, clip,height=3.3cm]{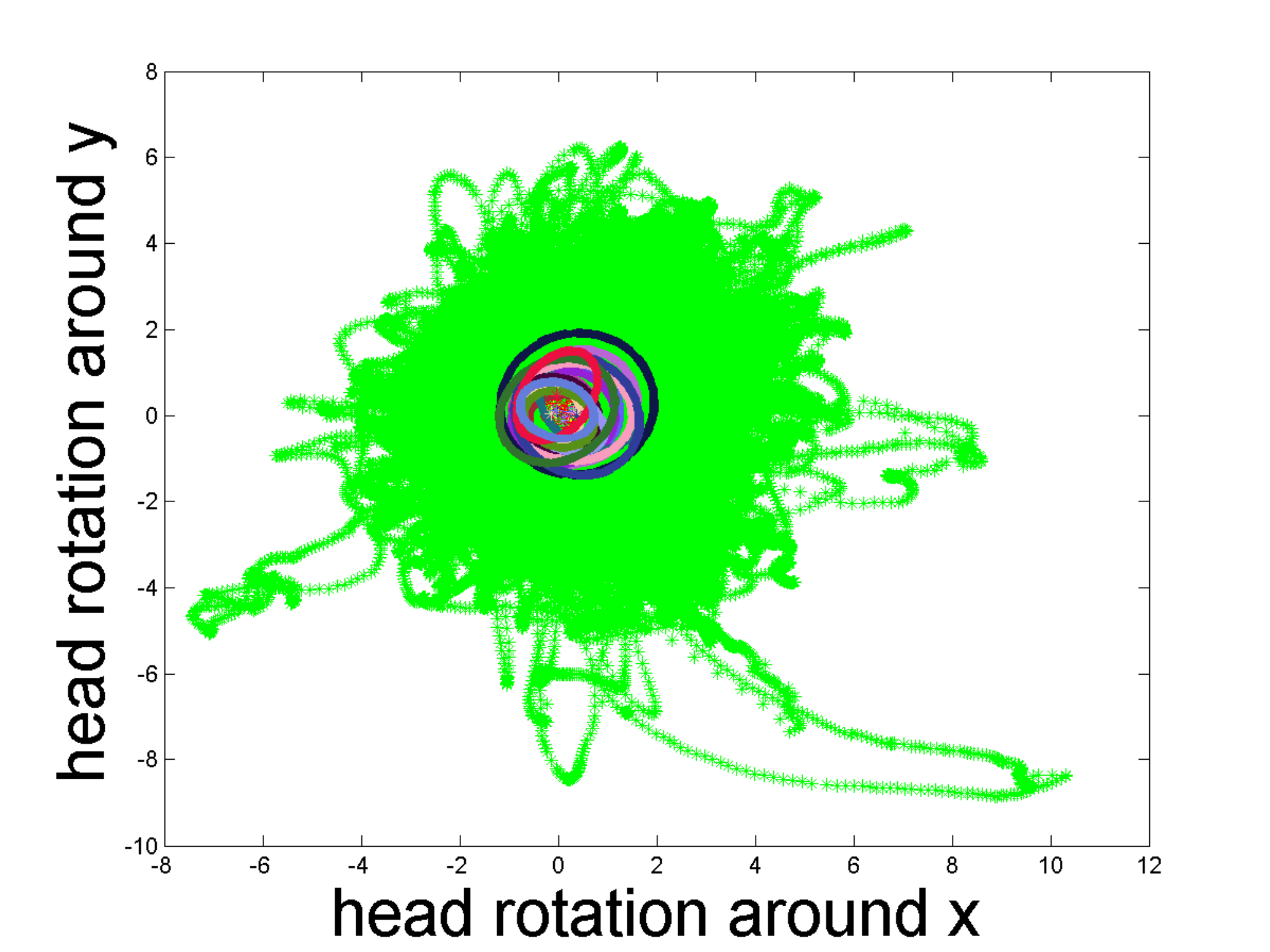}
                \label{fig:rand_init}
	}
        \subfigure[VQ based initialization]{
                \includegraphics[trim = 1cm 0cm 0.27cm 0cm, clip,height=3.3cm]{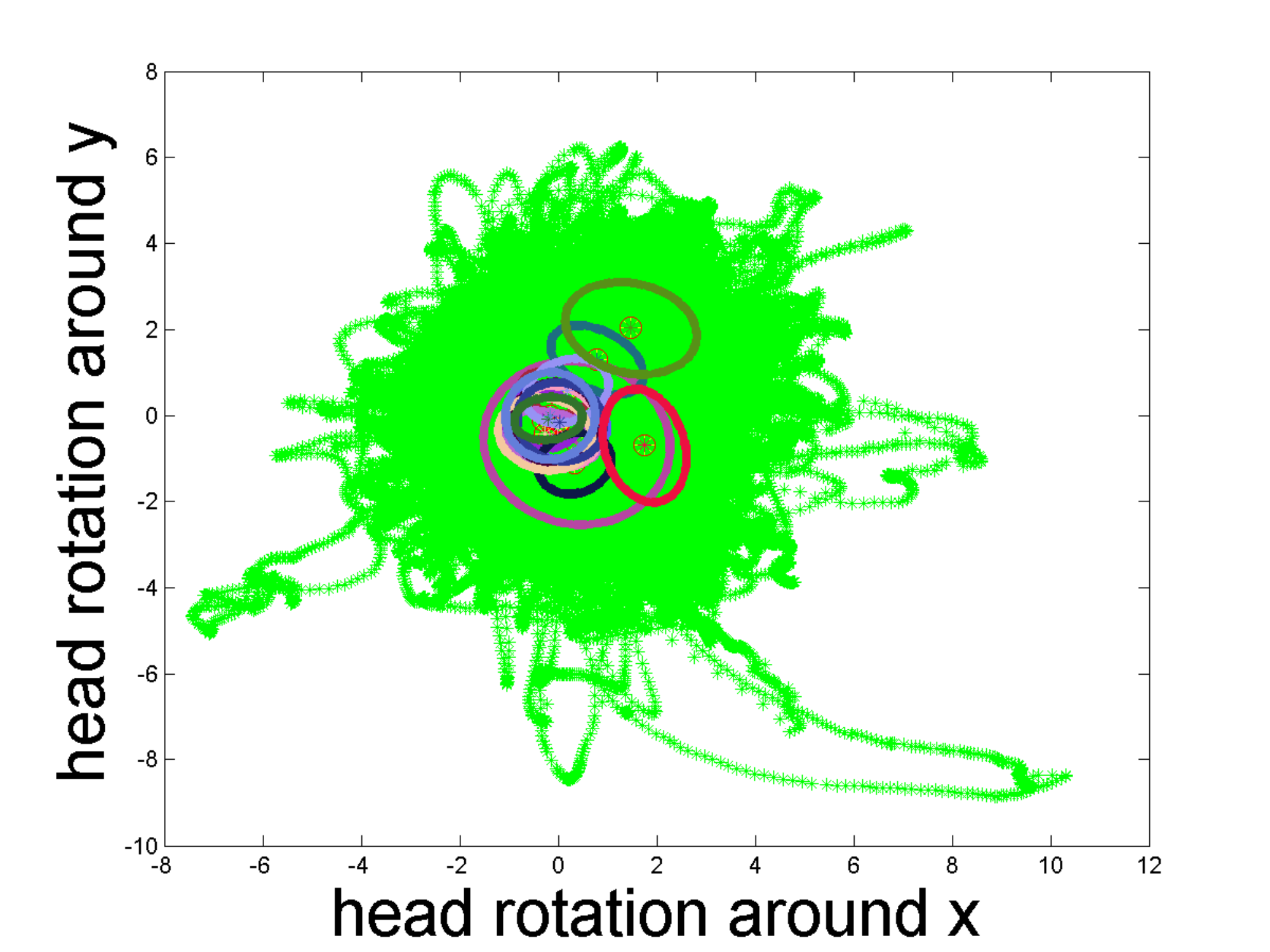}
                \label{fig:vq_init}
        }
\caption{Comparison of the means and covariance of the hidden states using random and VQ-based initializations. The VQ-based method increases the range of behaviors generated by the models.}
\label{fig:init}
\end{figure}

The parameters of the models are learned with conventional \emph{expectation maximization} (EM). Since EM finds local optimum, the initialization is very important. In Mariooryad and Busso \cite{Mariooryad_2012_2}, we randomly initialized the models. Since the generated behaviors correspond to the expected values given the speech features (Eq. \ref{eq:expectation}), the states may converge to the average position of the behaviors, reducing the range of behaviors generated by the model. Figure \ref{fig:rand_init} visualizes this problem. The figure shows the distribution of the original data for two head angles. Each ellipse represents the 16 states after training the models with random initialization. Each ellipse is centered at its mean vector and shaped according to its covariance matrix. The figure shows that all the clusters converge to the origin limiting the range of behaviors in the models (e.g., limited variability in the codebook for the speech-behavior space). To address this problem, we increase the representation of the initial states by using the \emph{Linde-Buzo-Gray vector quantization} (LBG-VQ) technique \cite{Linde_1980}. After splitting the data into clusters, we initialize the statistical properties of the states ($\mu^i$ and $\sigma^i$) using the results of the LBG-VQ. This approach leads to sparser states which increase the range of behaviors generated by the models. Figure \ref{fig:vq_init} shows the final 16 states achieved by the VQ-based initialization for two angles of the head position. The Figure shows that the VQ based initialization reaches a better representation of the data.

We smooth the trajectories generated by this network following the approach proposed by Busso \etal\cite{Busso_2007}. The method selects equidistant key-points. The value of the joint rotations in these key-points are transformed into their quaternion representation, where they are interpolated. The interpolation connects the key-points providing smooth transitions. We implement this method using 12 key-points per second for the hand motion, and 15 key-points per second for the head motion.

\section{Proposed Constrained Models}
\label{sec:CDBN}

This section describes the proposed model built upon the improved version of the DBN proposed by Mariooryad \etal \cite{Mariooryad_2012_2} described in Section \ref{sec:Baseline}. The key goal is to introduce constraints to generate meaningful behaviors. The constraints are either discourse functions or predefined prototypical gestures. The discourse function constraints bridge the gap between rule-based and data-driven system. The prototypical gesture constraints can serve as the behavior realizer in rule-based systems, capturing the intrinsic variability of each gesture, while preserving their temporal coupling with speech. 

\subsection{Adding Constraints to the DBN Model}
\label{ssec:cm1}

\begin{figure}
	\centering
\includegraphics[trim = 40mm 65mm 75mm 20mm, clip, width=6cm]{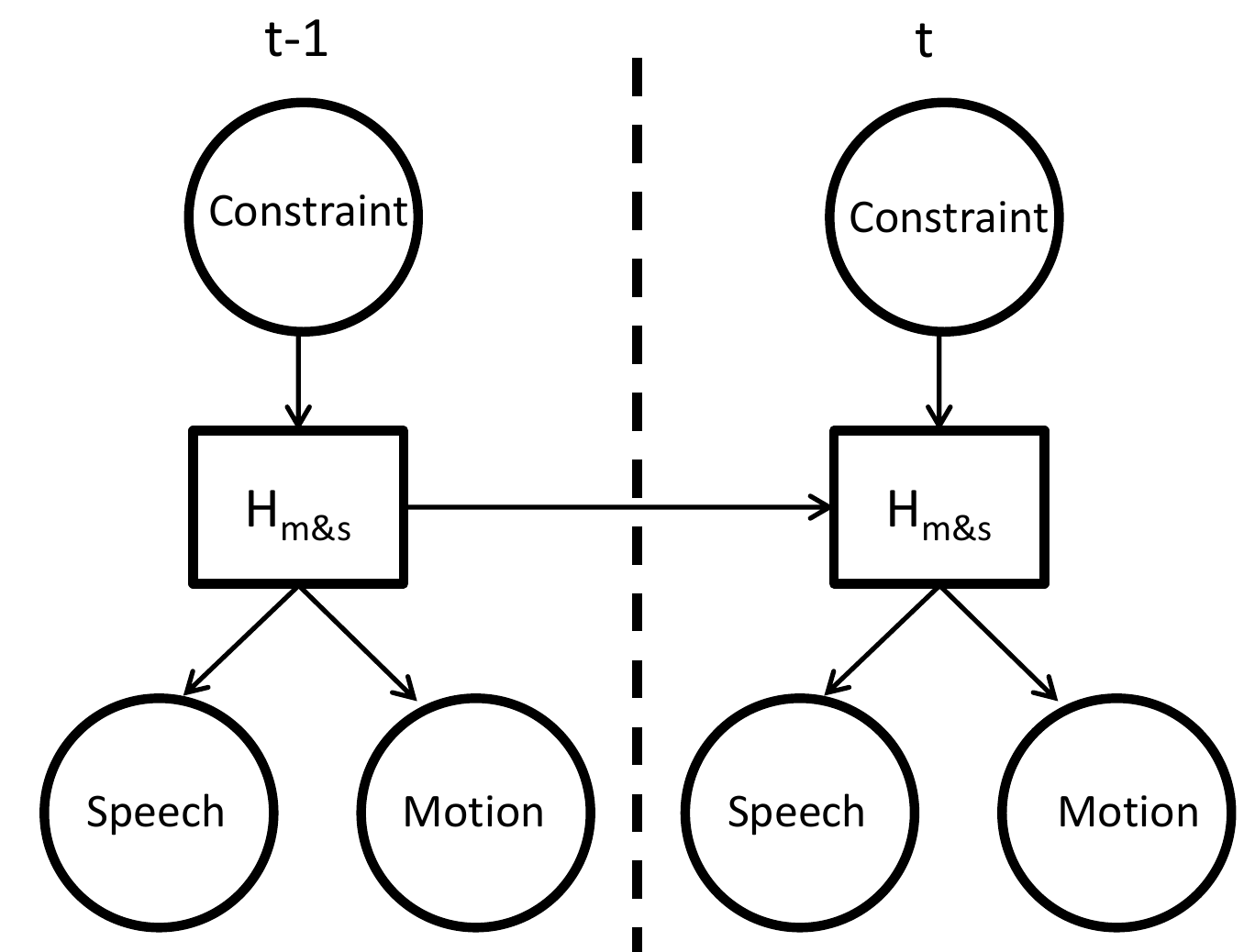}
\caption{Proposed framework that adds a constraint to generate meaningful data-driven behaviors.}
\label{fig:M1_}
\end{figure}

Figure  \ref{fig:M1_} illustrates the constrained model proposed in this study, which we refer to as  \emph{Constrained DBN} (CDBN). The key addition with respect to the baseline model is the node \emph{Constraint} which is introduced as a parent of the hidden state variable $H_{m\&s}$. With this additional node, the state variable is directly conditioned on the given constraint, affecting the relationship between gesture and speech. Effectively, this model has transition matrices, prior probabilities, and state prior probabilities for each constraint, learning the intrinsic characteristics of the gestures conditioned on the given constraint. 

This structure is different from the model proposed in our previous work, where the constraint was introduced as a child \cite{Sadoughi_2014}. By adding the constraint node as a parent we obtain the following advantages: (1) we separately  model  the prior probabilities of the constraints and their affect on the hidden states, (2) we handle constraint categories with unbalanced training data, and (3) we model a more reasonable cause-effect relationship between the variables.

The constraint added to the baseline model is a discrete observation node, representing the presence of a given constraint for each frame. We add the label \emph{other} when the constraint is not specified as an input.  Equation \ref{eq:trans} explicitly highlights that the transition probabilities $a_{kij}$ from the previous state to the current state depend on both the previous state and the current constraint: 

\begin{equation}
\begin{aligned}
a_{kij} = P\left[q_t = j|q_{t-1} = i, C_{t} = k\right],
\end{aligned}
\label{eq:trans}
\end{equation}

\noindent
where $q_t$ is the state at time $t$, $C_t$ is the constraint at time $t$, and $a_{kij}$ is the transition probability between the $i^{th}$ and $j^{th}$ state when the constraint is $k$. During synthesis, partial observation for this model includes \emph{Speech} and \emph{Constraint}. Equation \ref{eq:sampling_constrained1} defines the expected value for the node \emph{Motion} using partial inference. In this equation, $c_{1:t}$ represents the constraint sequence for the whole turn, meaning that $\gamma_{c_{1:t}}(i)$ depends not only on the $Speech_{1:t}$, but also on the $Constraint_{1:t}$.

\begin{equation}
\begin{aligned}
E\left[Motion|Speech, Constraint\right] =
\sum_{i=1}^n\mu_{h}^i\gamma_{c_{1:t}}(i)
\end{aligned}
\label{eq:sampling_constrained1}
\end{equation}

\subsection{Training Sparse Transition Matrices}
\label{ssec:SparseMatrix}

\begin{figure}
\centering
\includegraphics[trim = 60mm 70mm 60mm 10mm, clip, width=8cm]{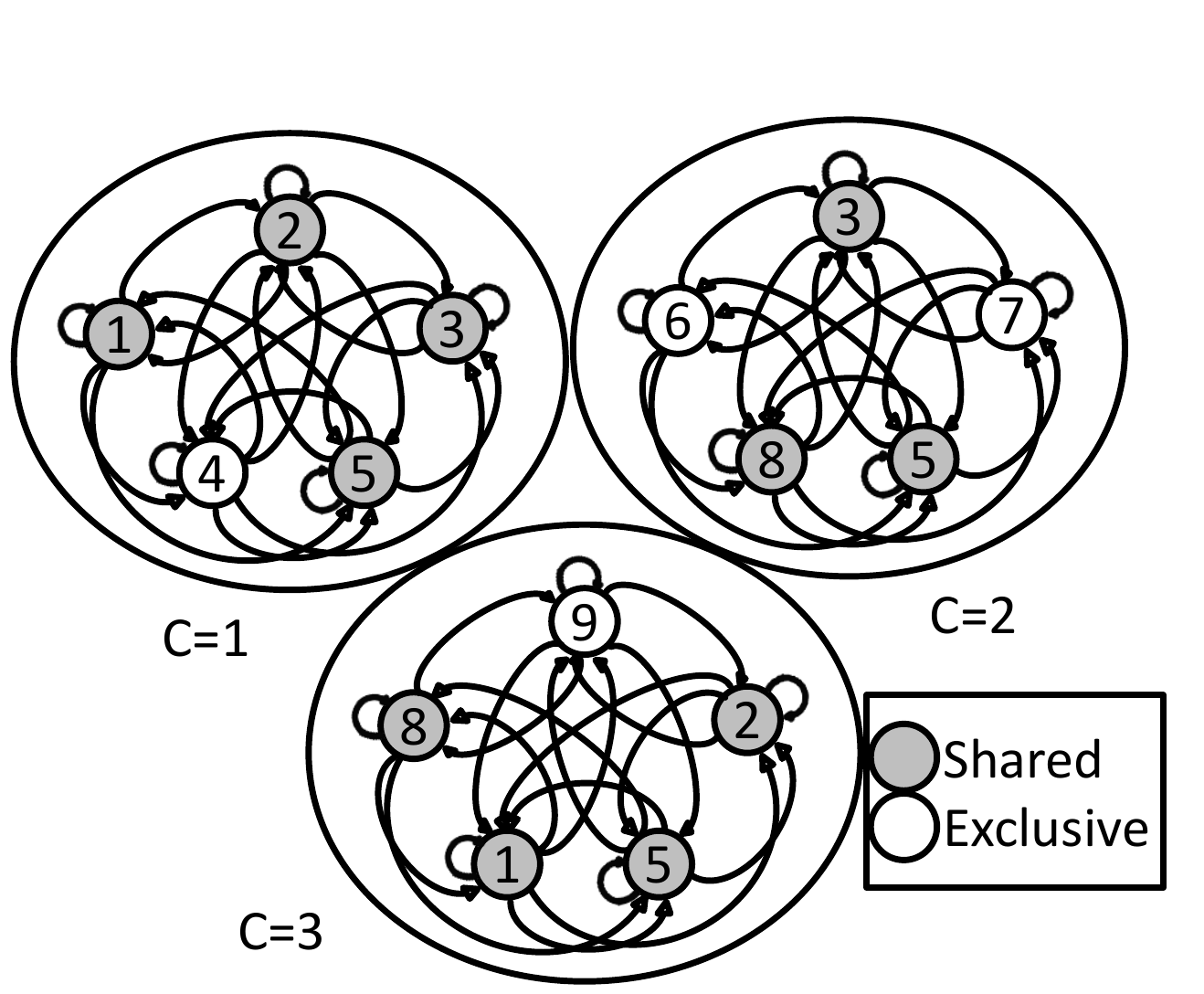}
\caption{The state graph of the model with sparse transition matrix for a model with 3 constraints and 5 states per constraint (C: Constraint).}
\label{fig:M2}
\end{figure}

The characteristic patterns associated with each constraint are captured by the constraint-dependent transition matrices ($a_{kij}$). If these transition probabilities are similar, the behaviors generated after imposing the constraints will also be similar, and the model will fail to generate the characteristic patterns analyzed in Section \ref{sec:Analysis}. As a result, we want to increase the differences in the transition probability assigned to each constraint. For this purpose, we propose a novel training approach to make the conditional transition matrices sparse. First, we create $N$ states per constraint in node $H_{m\&s}$, which are separately trained using the data associated with the given constraint (e.g., data annotated with either discourse functions or prototypical gestures). These $N$ states capture the characteristic patterns for each discourse function. If we have $K$ constraints, this step will generate $N \times K$ states. Using all these states is not practical since it unnecessarily increases the number of states in $H_{m\&s}$, and, therefore, the number of parameters. Furthermore, many of these states are redundant. Instead, we merge similar states, creating shared states and constraint specific states. We merge similar states using \emph{Kullback-Leibler divergence} (KLD). Since each state is a multivariate Gaussian distribution, we use Equation \ref{eq:kldmvg} to find similar states:

\begin {equation}
\begin{aligned}
	KL(P_j,Q_i) =\frac{1}{2}\left[\right.&\left .\mathrm{tr}(\Sigma_{q_i}^{-1}\Sigma_{p_j})-\log ( \Sigma_{q_i}^{-1}\Sigma_{p_j}) \right.\\
	&\left. -d + (\mu_{q_i}-\mu_{p_j})^T \Sigma_{q_i}^{-1} (\mu_{q_i}-\mu_{p_j})\right]
\end{aligned}
\label{eq:kldmvg}
\end{equation}

\noindent
where $P_j$ and $Q_j$ are the multivariate conditional Gaussian distribution for states $p_i$, and $q_j$, with covariance matrices $\Sigma_{p_j}$ and $\Sigma_{q_i}$, and mean vectors $\mu_{p_j}$ and $\mu_{q_i}$, and $d$ is the dimension of the Gaussian. First, we select all the states associated with a constraint. For each of them, we find the closest state from the states associated with other constraints, as determined by the KLD metric. If the difference is less than a threshold (empirically set to 1), we merge the states, becoming a shared state across constraints. We sequentially repeat this process for the states of each of the constraints. Finally, we create a new state which is shared between all the constraints to allow transition between the constraints. The resulting conditional transition matrices for each constraint is sparse, allowing only transitions between the $N$ states plus the additional state shared across constraints. This is the initialization phase for the model, and the parameters are refined afterward using EM. Figure \ref{fig:M2} gives an illustration of the states for a model with 3 constraints and 5 states per constraint. States 1, 2, 3, 5 and 8 are shared across more than one constraint, and states 4, 6, 7, and 9 are exclusive.  

To illustrate the importance of these sparse transition matrices, we compare the transition matrices when (1) the $N$ states are shared across constraints, and (2) the $N$ states per constraint are defined using the proposed approach. We estimate these transition matrices for \emph{head shakes}, \emph{head nods}, and \emph{other}, using $N=8$. The average of the $L_\infty$ distances between the transition matrices conditioned on \emph{Head Nods} (A), and \emph{Head Shakes} (B) are 0.018 for option 1 (shared states) and 0.96 for option 2 (sparse matrices). This result shows that the training approach is more successful at capturing the differences between different constraints. Therefore, we rely on this approach for training the CDBN models.

\begin{equation}
d_\infty(A,B)=\max_{1\leq i \leq N} \max_{1\leq j \leq N} |a_{ij} - b_{ij}|
\label{eq:dist_mat}
\end{equation}

\begin{figure}
\centering
	\includegraphics[trim = 10mm 10mm 17mm 0mm, clip, width=2cm]{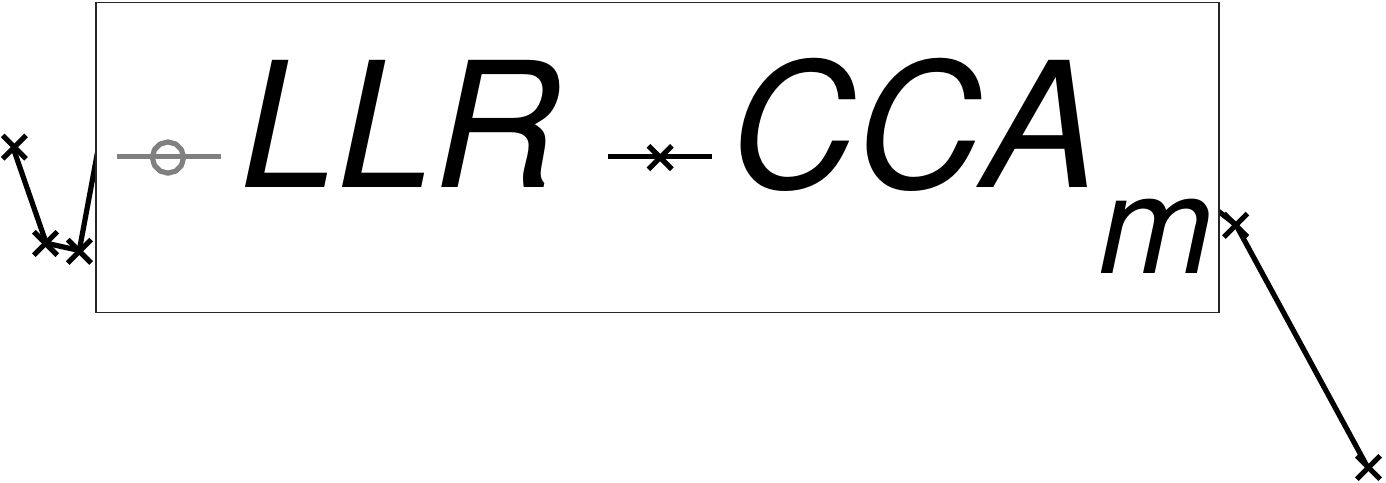}\\
\subfigure[CDBN, constrained on Discourse for head]{
\includegraphics[trim = 0mm 0mm 0mm 8mm, clip, width=6cm]{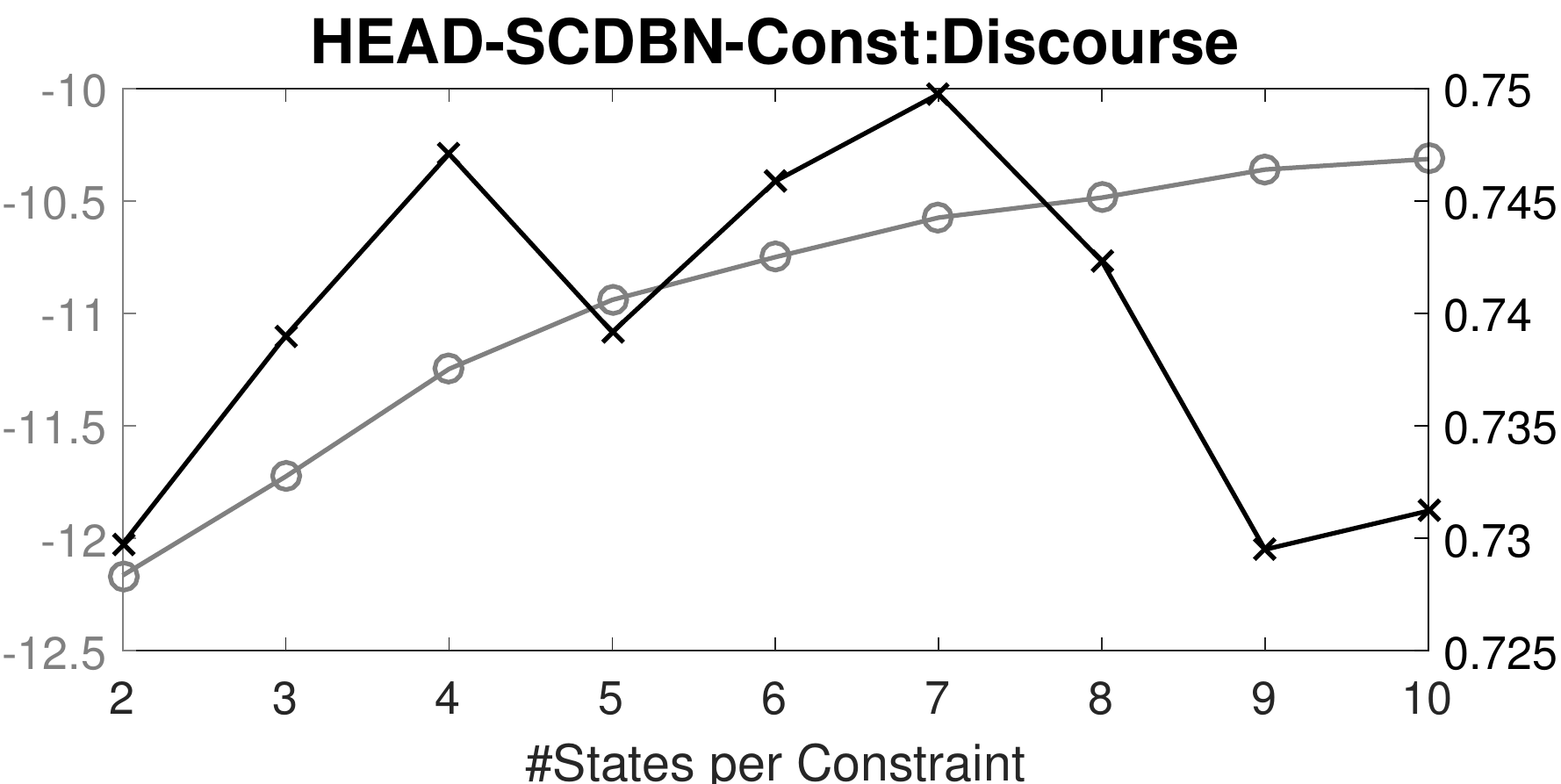}
}
\subfigure[CDBN, constrained on Discourse for hand]{
\includegraphics[trim = 0mm 0mm 0mm 8mm, clip, width=6cm]{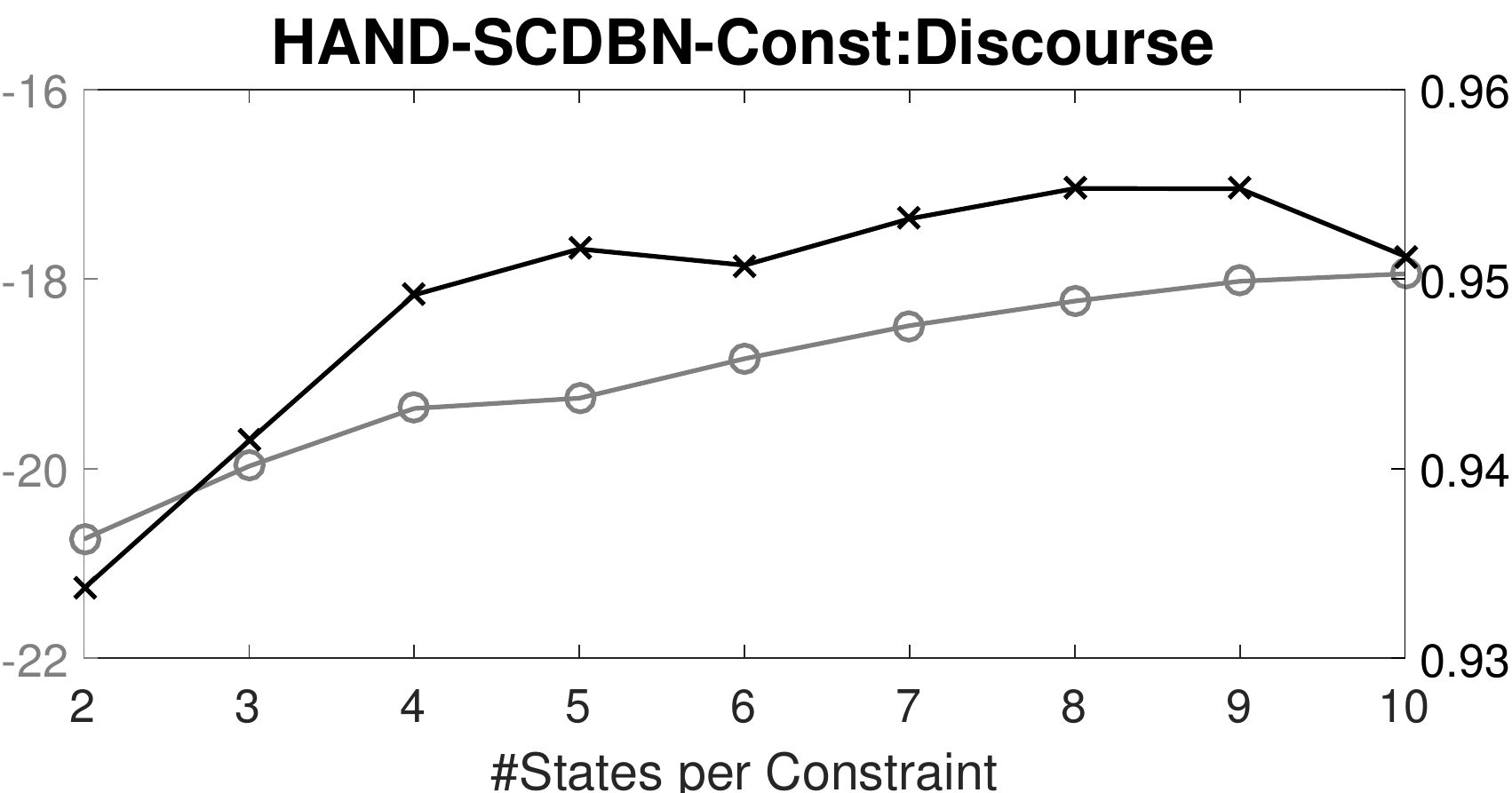}
}
\caption{Changes of LLR and CCA as we increase the number of states for the model constrained on discourse functions for hand and head movements.}
	\label{fig:nbst}
\end{figure}

\section{Experiments \& Results}
\label{sec:results}

This section reports the experiments and the results of the evaluation of the proposed models. The baseline model is the improved framework proposed by Mariooryad and Busso \cite{Mariooryad_2012_2} (Sec. \ref{sec:Baseline}). We compare the models using objective and subjective metrics. We separately train the models for head and hand, since the gesture constraints do not  necessarily coincide. The evaluation relies on ten-fold cross-validation to maximize the use of the corpus. 

\subsection{Optimization of Number of States}
\label{ssec:NumberofStates}

Before training the models, we need to determine the number of states. Optimizing this parameter is computationally expensive since the CDBN models need to be trained multiple times as we vary $N$, repeating the approach for each fold in the cross-validation process. Therefore, we simplify the evaluation by setting one of the ten partitions as a validation set. We use the other nine partitions to train the models. This process is conducted once, using the optimal parameters for the rest of the evaluation.

We use two objective metrics. The first metric is the average \emph{canonical correlation analysis} (CCA) between the original trajectory of the behaviors and the synthesized movements ($CAA_m$, where $m$ is either hand or head motion). CCA projects two multidimensional data into a common space where their correlations are maximized. The value range between 0 and 1, where 1 implies perfect correlation and 0 no correlation between the variables. We estimate $CCA_{m}$ per turn, reporting the average results. The second metric is the average \emph{log likelihood rate} (LLR) of the model ($\frac{\mathrm{log} P(O|\lambda)}{T}$), where $T$ is the number of frames, and $P(O|\lambda)$ is the observation probability given the model parameter $\lambda$. We determine the number of state such that the $CCA_{m}$ and the LLR of the model are both high.

We separately estimate the number of parameters for each model (baseline models for head and hand, CDBN models for head and hand with discourse and gesture constraints). Figure \ref{fig:nbst} shows an example of the changes observed for LLR and $CCA_{m}$ in the validation set for the CDBN model for head and hand motions constrained on discourse functions. This figure shows that the $CCA_{m}$ and LLR values start to saturate when the number of states is $N$=7 for head and $N$=8 for hand. We obtain similar figures for other cases, not reported in the paper, setting the optimal value for $n$. The row $\#States/Const$ in Tables \ref{tab:res-head} and \ref{tab:res-hand} provides the number of states per constraint (e.g., $N$) for all the conditions. We use these parameters for the rest of objective and subjective evaluations

\subsection{Objective Evaluation}
\label{ssec:Objective}

For the objective evaluation, we use ten-fold cross-validation approach. We avoid using the partition used for validation in the test set. Therefore, we only consider nine folds where the test set  in each cross validation is one of the remaining nine partitions. After selecting the test set, we form the training set with the other nine partitions, adding the partition used for validation.

We separately evaluate the models constrained on discourse functions (CDBN-Dis) and prototypical gestures (CDBN-Ges) by comparing the generated trajectories with the baseline model (Sec. \ref{sec:Baseline}). We evaluate the generated movements in terms of the CCA between the original and generated motion sequences ($CCA_m$), and the CCA between the generated motion sequences and the speech sequence ($CCA_{ms}$). We also use the KLD. The $KLD(p||q)$ measures the amount of information lost when distribution $q$ is used to represent distribution $p$. We evaluate the KLD between the synthesized movements ($q$) and the original movements ($p$). Ideally, this value should be as small as possible, indicating that the generated movement sequences have similar distributions as the original motion sequences.

\begin{table}[t]
	\centering
	\caption{Objective evaluations on the models trained for head movements. The table lists the number of states. Asterisks denote significant differences for the corresponding metrics with respect to the baseline model (**: $p<0.001$).}
	\begin{tabular}{l c | c| c  c }
		\hline
		\multicolumn{2}{c|}{Region} &\multicolumn{3}{c}{Head Movements}\\\hline\hline
		\multicolumn{2}{l|}{Model} & Baseline & CDBN-Dis &  CDBN-Ges\\\hline
		\multicolumn{2}{l|}{\#States} & 7 & 28.1 &  22.2 \\
		\multicolumn{2}{l|}{\#States/Const} & 7 & 7 & 8  \\
		\multicolumn{2}{l|}{\#Params} &  300 & 1000.7  & 786.0\\
		\hline\hline
		\multirow{2}{*}{CCA$_m$}		& M & 0.680 &   0.691  &  0.677  \\
								& SD &  0.226 &  0.222  & 0.230  \\
		\hline
		\multirow{2}{*}{CCA$_{ms}$}	& M & 0.823  &  0.770**  & 0.760**  \\
								&SD & 0.135 &   0.186 &   0.187   \\
		\hline
		\multicolumn{2}{l|}{KLD} & 7.593  &  2.718  &    2.957\\
		\hline
	\end{tabular}
	
	\label{tab:res-head}
\end{table}

\begin{table}[t]
	\centering
	\caption{Objective evaluations on the models trained for hand movements. The table lists the number of states. Asterisks denote significant differences for the corresponding metrics with respect to the baseline model (*: $p<0.05$).}
		\begin{tabular}{l c | c | c c }
		
		\hline
		\multicolumn{2}{c|}{Region} &\multicolumn{3}{c}{Hand Movements}\\\hline\hline
		\multicolumn{2}{l|}{Model} & Baseline &CDBN-Dis& CDBN-Ges\\\hline
		\multicolumn{2}{l|}{\#States} &  12 & 40 &37\\
		\multicolumn{2}{l|}{\#States/Const} & 12 & 8 & 9\\
		\multicolumn{2}{l|}{\#Params} &   1247 & 3356.0 & 3130.0\\
		\hline\hline
		\multirow{2}{*}{CCA$_m$}& M &   0.933  &  0.942*    &   0.936 \\
							& SD &  0.100   &  0.086      & 0.096\\
		\hline
		\multirow{2}{*}{CCA$_{ms}$} 	& M &   0.927 &   0.924     &  0.923\\
								&SD &   0.079   & 0.083  &    0.086\\
		\hline
		\multicolumn{2}{l|}{KLD} 			&   1.346 &   1.176 & 0.827\\
		\hline
	\end{tabular}
	\label{tab:res-hand}
\end{table}

As a reference, the CCA between the original head motion and speech is $\rho=0.694$, and the CCA between the original hand motion and speech is $\rho=0.847$. These high correlation values show the strong coupling between the joint movements and the prosodic features.  Tables \ref{tab:res-head} and \ref{tab:res-hand} give the results for the synthesized head and head movements, respectively. The results show that the constrained model on discourse function, achieves higher $CCA_{m}$ than the baseline model for the hand region (t-test: $p=0.0269$) . The results also indicate that the constrained models for head region give lower $CCA_{ms}$ than the baseline model (t-test: $p<1e^{-11}$). While the constrained models reduce the coupling between the generated trajectory and speech, their values are still higher than the coupling between original head movement and speech (e.g., $\rho=0.694$). As demonstrated by the subjective evaluation, the movements are more natural and appropriated when we constrain the models with discourse functions. The results for the KLD shows improvements on all the constrained models compared with the baseline models. The distributions of the generated behaviors are closer to the distributions of the original trajectories, compared to the baseline model.

\subsection{Subjective Evaluations}
\label{ssec:subjective}

This section reports the subjective evaluations of the behaviors generated with the proposed models. We use the Smartbody toolkit \cite{Thiebaux_2008} for rendering the movements, where the only variable that we control is the hand gesture and head motion. Everything else is kept consistent across conditions (e.g., facial expressions). We separately evaluate the models constrained on either discourse functions or prototypical gestures. The train and test partitions are the same as the ones described in Section \ref{ssec:Objective}.  

The first part of the subjective evaluation is when we constrain the models on the discourse functions (Sec. \ref{sec:annotation-of-discourse-function}). We evaluate the perceived appropriateness and naturalness of the movements generated by the baseline model and CDBN model constrained on the discourse functions. For each constraint, we randomly selected 10 segments labeled with the corresponding discourse function. To provide enough context, we include the speaking turn preceding the selected turn. The animation is idle when the CA is listening to the other speaker (the MSP-AVATAR corpus consists of dyadic scenarios).  

\begin{figure}
	\centering
	\includegraphics[trim = 21mm 0mm 31mm 0mm, clip, width=6cm]{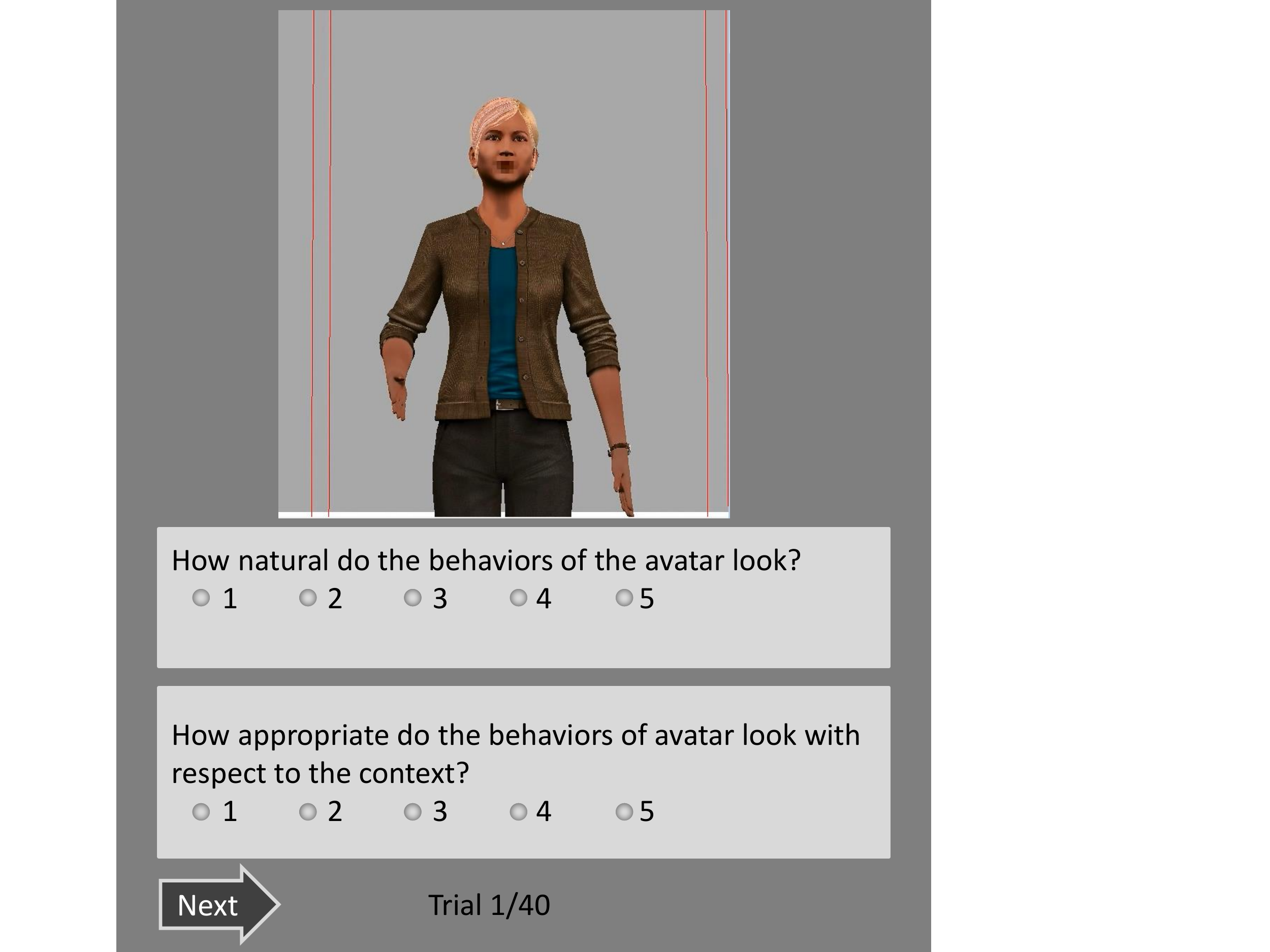}
	\caption{The interface for rating the animations based on appropriateness and naturalness using five-point Likert-like scales. The questions are displayed after the video is played.}
	\label{fig:mturk2}
\end{figure}

We use \emph{Amazon mechanical turk} (AMT) for the perceptual evaluations, using the interface shown in Figure \ref{fig:mturk2}. We display the questions after the video is played to assure that the evaluators do not answer the questions before the video is played. We randomize the order of the videos for each evaluator. We only allow workers from the United States with overall acceptance rate of more than 80\%.

We evaluate $120$ segments ($4$ discourse functions $\times 3$  conditions  $\times 10$  videos) animated using (1) the original motion capture data (\emph{Original}), (2) the baseline model (\emph{baseline}), and (3) the CDBN models constrained on discourse functions (\emph{CDBN}). Each video was annotated by five different evaluators. We asked the subjects to rate the animations in terms of naturalness and appropriateness of the movements using a five-point Likert-like scale (Fig. \ref{fig:mturk2}). In total, we have 15 evaluators, where nine are females and six are males (average age is 31.1).

\begin{figure}
	\centering
	\subfigure[Appropriateness-All Constraint]{
		\includegraphics[trim = 0mm 0mm 10mm 0mm, clip, height=4.3cm]{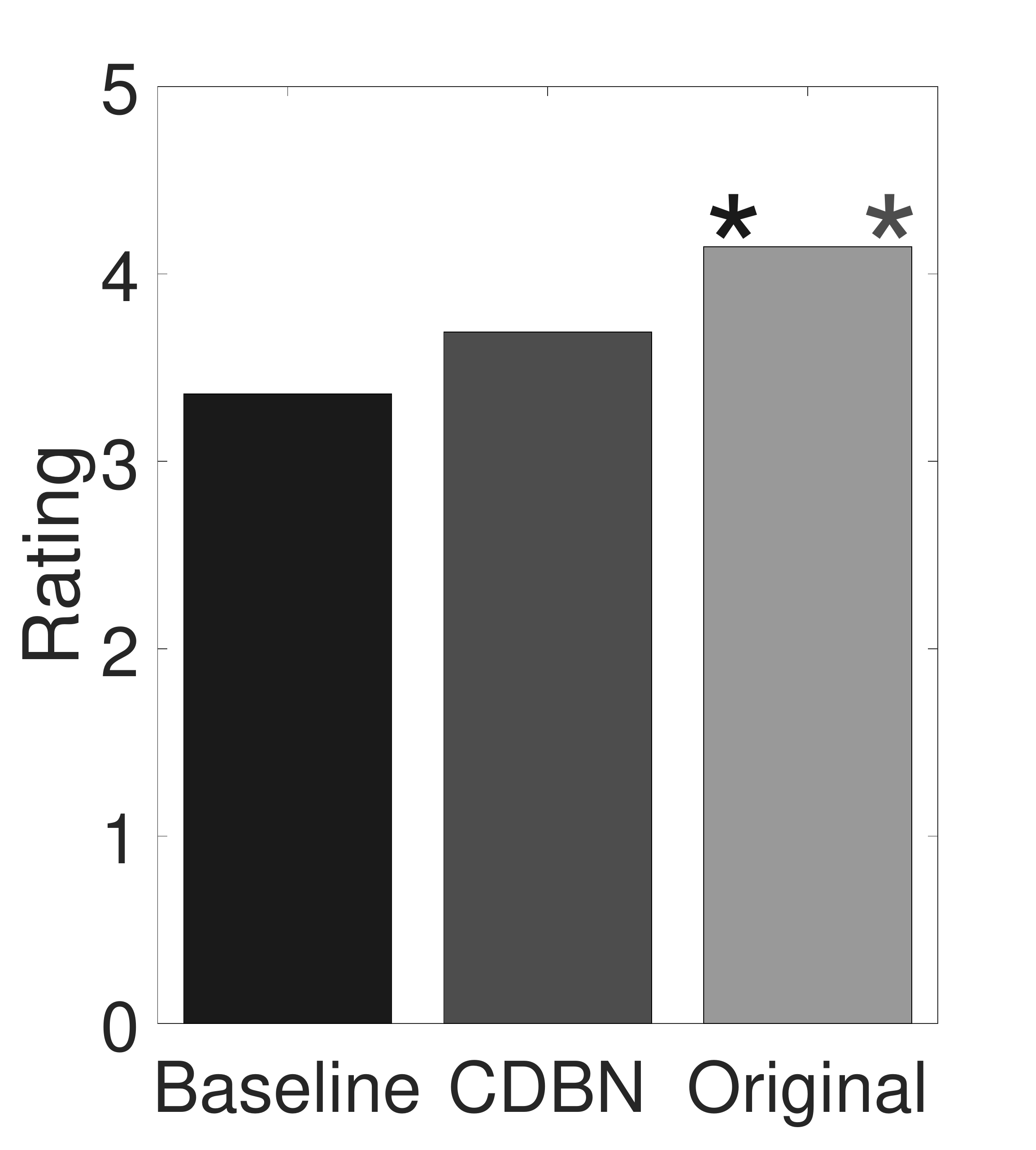}
		\label{fig:rating_allc-a}
	}
	\subfigure[Naturalness-All Constraint]{
		\centering
		\includegraphics[trim = 0mm 0mm 10mm 0mm, clip, height=4.3cm]{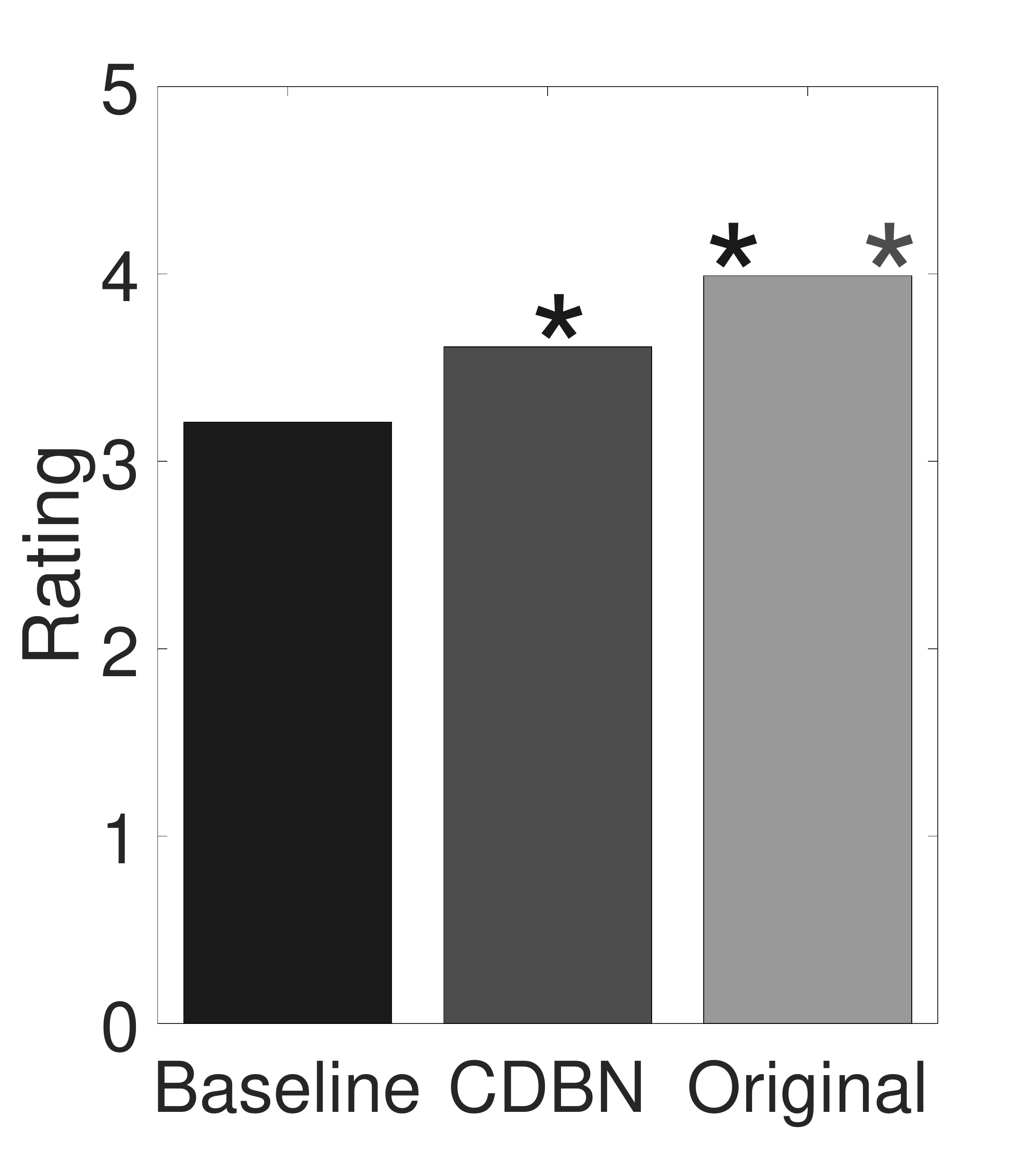}
		\label{fig:rating_allc-n}
	}
	\caption{Results of the perceptual evaluations when the constraints are discourse functions. The bars represent the means per condition. Asterisks represent statistically higher values with respect to the bar indicated by the color of the asterisk ($p$-value$<$0.05).}
	\label{fig:rating1}
\end{figure}

Figures \ref{fig:rating_allc-a} and \ref{fig:rating_allc-n} give the average of the ratings given by the evaluators for appropriateness and naturalness, respectively. The Cronbach's alpha between the annotators is $\alpha=$0.48. The Kruskal-Wallis test shows that the videos synthesized by the three conditions are different ($p<1e^{-10}$). The pairwise comparisons of the results are denoted in the figures with a color coded asterisks.  The color of the asterisk indicates that the given condition is statistically higher than the condition associated with the bar with the given color (we assert performance at $p$-value$<$0.05). The pairwise comparison of the results shows that the original motion capture recordings are perceived as more natural and appropriate than the animations synthesized by both models ($p<0.001$). However, the CDBN models are perceived with higher level of appropriateness and naturalness than the baseline models. The difference is statistically significant for naturalness ($p<0.01$). 

\begin{figure}
	\centering
	\includegraphics[trim = 40mm 75mm 25mm 10mm, clip, width=4cm]{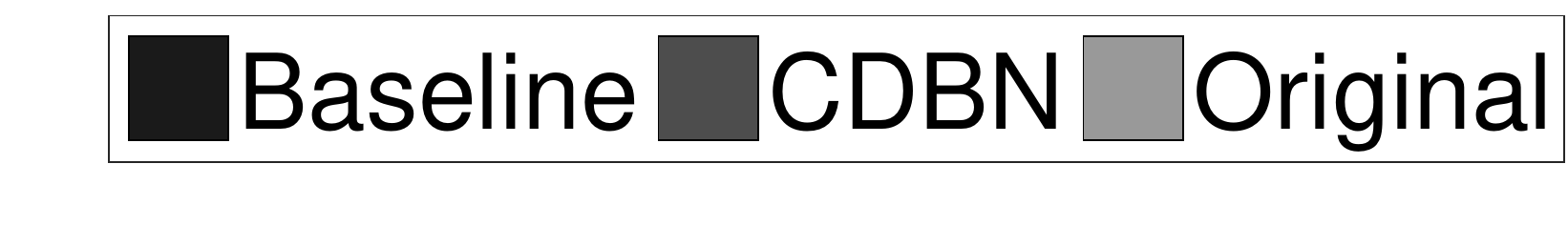} \vspace{-0.3cm}
	
	\subfigure[Appropriateness-Per Constraint]{
		\includegraphics[trim = 0mm 0mm 10mm 0mm, clip, height=4.4cm]{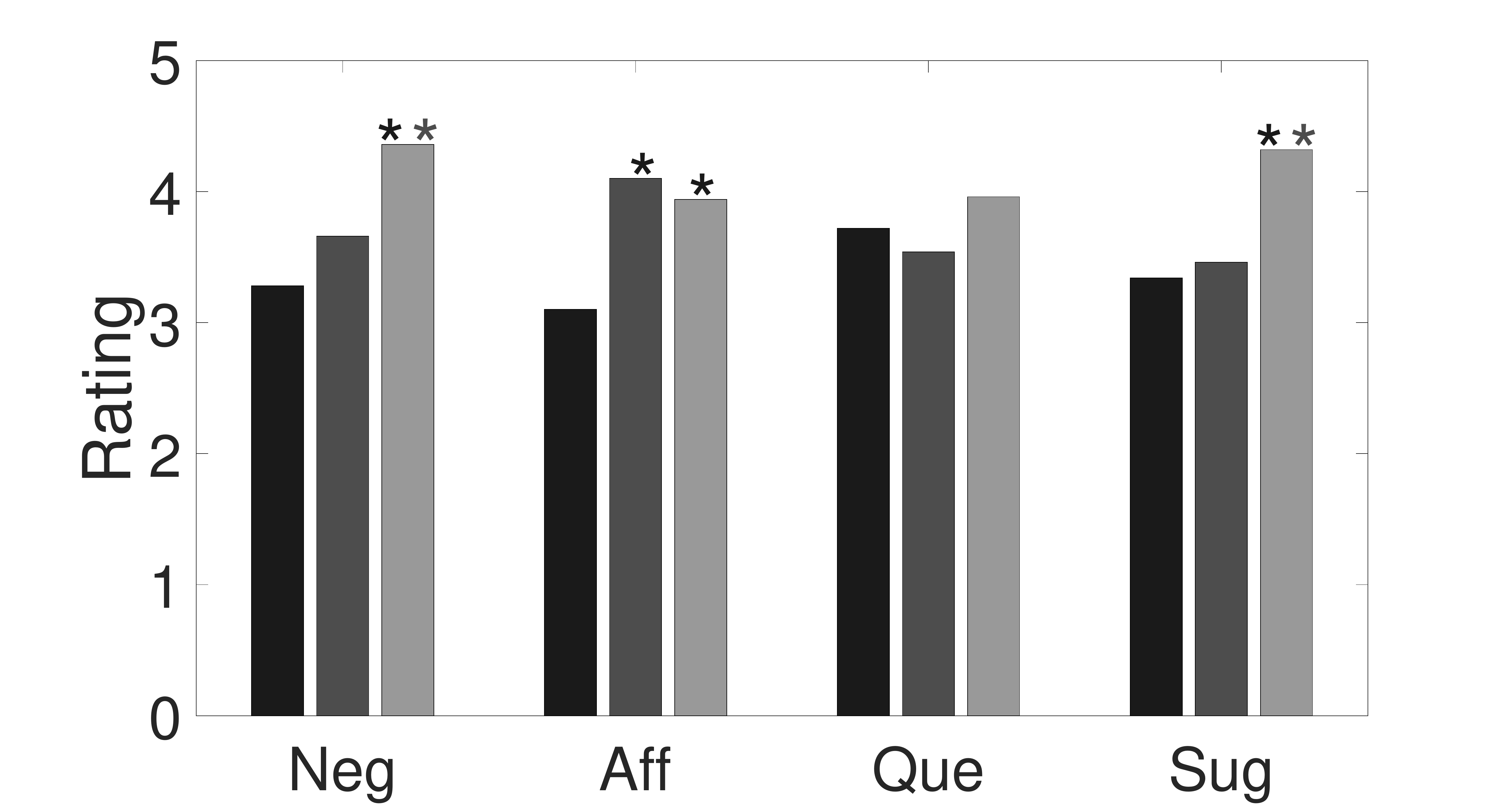}
		\label{fig:rating_perc-s}
	}     
	\subfigure[Naturalness-Per Constraint]{
		\includegraphics[trim = 0mm 0mm 10mm 0mm, clip, height=4.4cm]{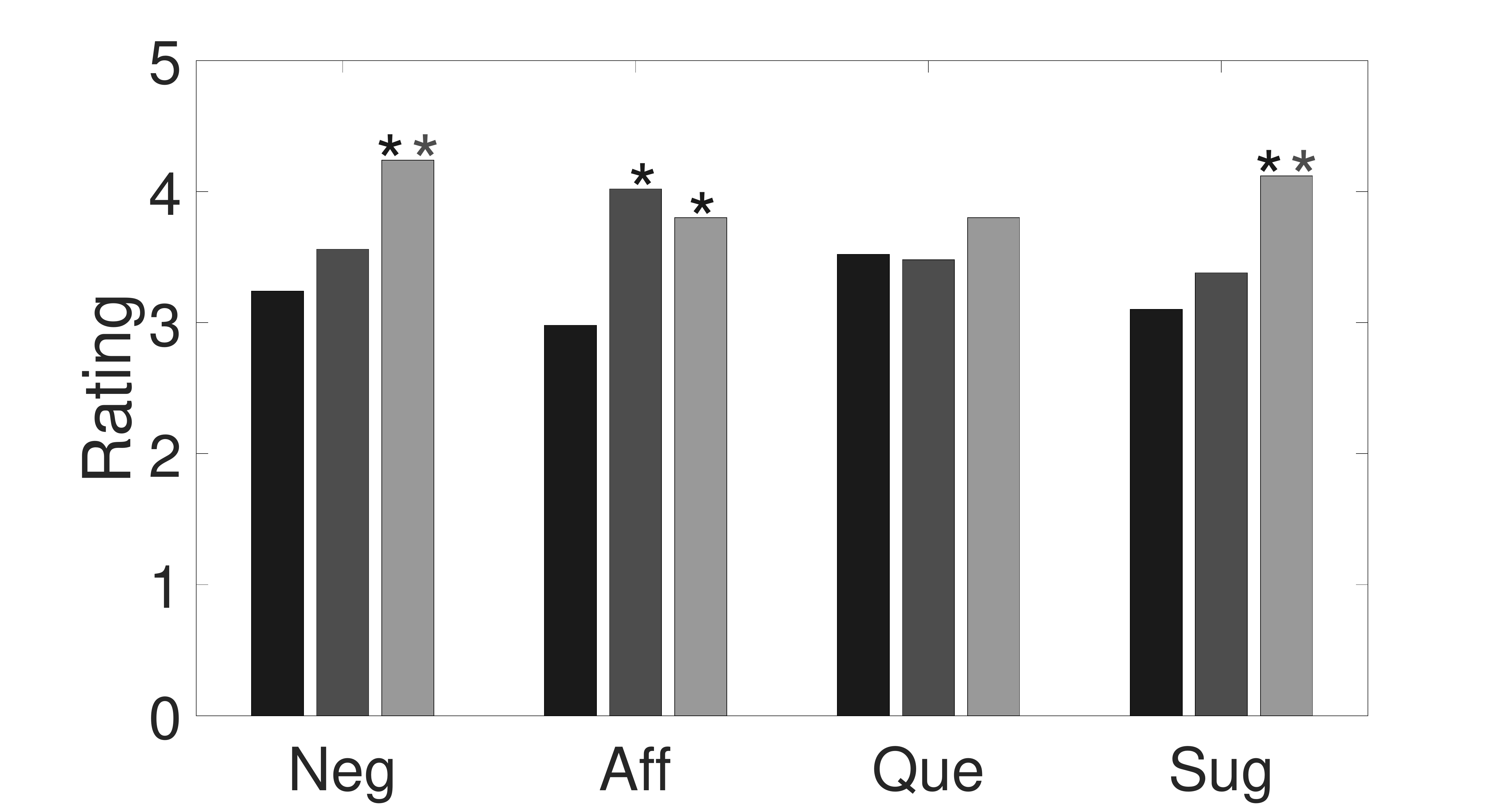}
		\label{fig:rating_perc-n}
	}
	\caption{Results of the perceptual evaluations per discourse function. The bars represent the means per condition. Asterisks represent statistically higher values with respect to the bar indicated by the color of the asterisk ($p$-value$<$0.05).}
	\label{fig:rating2}
\end{figure}

We also analyze the performance of the constrained model per discourse functions. Figure \ref{fig:rating2} gives the results. With the exception of \emph{questions}, the CDBN model improves the perception of naturalness and appropriateness over the baseline models. The differences are statistically different for \emph{affirmation} ($p<0.001$), where the values are slightly higher than the videos rendered with the original sequences. The consistency in the results reveal that the proposed models and training strategy can effectively capture the range of behaviors characteristic of the given constraint.

The second part of the subjective evaluation is when we constrain the models on the prototypical behaviors (Sec. \ref{ssec:Prototypical}). We synthesize 60 segments per gesture (i.e., where the constraint is the target gesture). We randomly choose these segments from the fully annotated sessions. We find the accuracy per gesture by watching the animations generated for these segments, where a success is considered when the generated behavior matches the target gesture. Table \ref{tab:res_const_ges} gives the accuracy for different head and hand gestures. The generated head gestures for \emph{Nod} and \emph{Shake} match the target gesture more than 80\% of times. This high accuracy demonstrates the benefits of the proposed constrained models. For hand movements, the gesture \emph{So-What} has the highest accuracy (85\%). The accuracy for \emph{To-Fro} and \emph{Regress} is not as high. We hypothesize that the accuracy for \emph{Regress} may be due to the high variability observed in the training samples. For \emph{To-Fro}, the result may be related to the lower precision of the samples retrieved for this prototypical behavior (Table \ref{tab:res_det}). The results from these evaluations are encouraging, suggesting that there is room for improvements. 

\begin{table}
	\centering
	\caption{Accuracy of the synthesized gestures using the CDBN framework when we constrain the models on prototypical gestures.}
	\begin{tabular}{c l| c}
		\hline
		\multirow{2}{*}{Region}& \multirow{2}{*}{Behavior} & Accuracy\\
        &&[\%]\\
		\hline
		\hline
		\multirow{2}{*}{HEAD} 
		& Nod  & 81.7\\
		& Shake & 80.0\\
		\hline
		\multirow{3}{*}{HAND} 
		& To-Fro & 61.7\\
		& So-What & 85.0\\
		& Regress & 55.0\\
		\hline
	\end{tabular}
	\label{tab:res_const_ges}
    \vspace{-0.2cm}
\end{table}

\section{Conclusions}

This paper explored the idea of introducing constraints in speech-driven models to generate behaviors with meaning that are timely coupled with speech. We evaluated a unified model with two types of constraints for hand and head movements: discourse functions and prototypical gestures. We incorporated discourse functions into the speech-driven framework to capture the characteristics behaviors associated with each of the four classes considered in the study (negations, affirmations, questions and suggestions). As demonstrated  by the analysis, individuals displayed characteristic patterns for these discourse functions, which our model aims to capture (e.g., generating meaningful gestures when people are asking questions). Likewise, we constrained the models with predefined prototypical gestures for head (shake, nod) and hand (so-what, to-fro, regress) gestures.  This model can be used by a rule-based system as a behavior realizer. The proposed approach not only creates the appropriate behavior, but also captures the temporal coupling between speech and the synthesized movement, which is not easily achieved with only rule-based systems.

The proposed framework is built upon the DBN models proposed by Mariooryad and Busso \cite{Mariooryad_2012_2}, providing three important contributions to effectively constrain the models on the underlying discourse function or prototypical gesture. First, we introduce a variable that constrains the state configuration between speech and gestures, capturing the cause-effect relation of the gesture production. Second, we proposed a better initialization of the models using vector quantization. The proposed training approach effectively increases the range of the movements generated by the model. Third, we introduced shared and exclusive states for each of the constraints, creating sparse transition probability matrices. Some of the states are shared between constraints, while other are exclusively associated with a constraint. This approach effectively captures the differences in the behaviors across constraints. 

The results from the objective and subjective evaluations demonstrated the benefits of the proposed approach. The results of the perceptual evaluation showed significant improvement for the constrained model over the unconstrained baseline model, especially for \emph{affirmation}. The results for prototypical gestures also revealed the potential of the proposed work. The head gestures synthesized by the constrained model generated the target gesture with 80\% accuracy. The hand gestures generated by the constrained model showed 85\% accuracy for \emph{so-what}. For \emph{to-fro}, and \emph{regress} the accuracies are lower. 

The study opens interesting opportunities to increase the role of data-driven models in CAs. For example, the proposed approach can be combined with rules driven from natural recordings \cite{Chiu_2015} to create meaningful and naturalistic gestures. One limitation of the approach is that it requires speech. We are exploring training schemes to extend the models by driving the behaviors using synthetic speech \cite{Sadoughi_2016,Sadoughi_201x2}. If we can solve the challenges in using synthetic speech instead of natural speech, we can increase the range of CA applications for data-driven models. With the transcription, we can also infer discourse functions using automatic algorithms. Dialog acts are semantic tags which can be retrieved from the text using supervised classifiers. These tags can then be translated into discourse functions, resulting in an autonomous meaningful behavior generator. Finally, we can address the lower performance for prototypical hand gestures by adding more data, capturing intrinsic variability between people, and by using more powerful frameworks. Advances in deep learning, in particular, offer appealing alternatives for this task \cite{Haag_2016, Lan_2016, Sadoughi_2017}.

\section*{Acknowledgment}
The authors would like to thank Sunghyun Park, Philippa Shoemark, and Louis-Philippe Morency for sharing the OCTAB interface. This work was funded by National Science Foundation grants IIS:1352950 and IIS:1718944.


\ifCLASSOPTIONcaptionsoff
  \newpage
\fi



%
\bibliographystyle{abbrv}
\bibliography{reference}

\end{document}